\documentclass[fleqn,11pt]{paper}

\usepackage[utf8]{inputenc}
\usepackage[T1]{fontenc}

\usepackage{authblk}

\usepackage[dvipsnames]{xcolor}
\usepackage[colorlinks=true, allcolors=Blue]{hyperref}

\usepackage[a4paper,margin=2.5cm]{geometry}
\usepackage{lmodern}

\usepackage{placeins}
\renewcommand{\arraystretch}{1.3} %stretches array slightly to make it more legible

\usepackage{colortbl} %Shading to highlight in array
\usepackage{booktabs, multirow, makecell}
\newcolumntype{g}{>{\columncolor{lightgray}}c}

\usepackage{cite}
\usepackage[sort&compress,numbers]{natbib}
\usepackage[nottoc]{tocbibind}
\newcommand*{\doi}[1]{\href{https://doi.org/#1}{DOI: #1}}

\usepackage{mathtools,amsmath, amssymb,amsthm}
\usepackage{tensor,mathrsfs,upgreek,stmaryrd}
\let\t\tensor
\allowdisplaybreaks

\usepackage{cancel}

\usepackage{xfrac}
\def\half{\tfrac{1}{2}}
\def\ihalf{\tfrac{\ii}{2}}

\def\quarter{\tfrac{1}{4}}

\DeclareFontFamily{OT1}{rsfs}{}
\DeclareFontShape{OT1}{rsfs}{CGNPm}{n}{ <-7> rsfs5 <7-10> rsfs7 <10-> rsfs10}{}
\DeclareMathAlphabet{\mycal}{OT1}{rsfs}{CGNPm}{n}

\numberwithin{equation}{section}
\numberwithin{figure}{section}
\numberwithin{table}{section}

\newcommand{\bx}{{\bar{x}}}
\newcommand{\by}{{\bar{y}}}
\newcommand{\bz}{{\bar{z}}}
\newcommand{\bA}{\bar{A}}
\newcommand{\bg}{\bar{g}}

\newcommand{\wepsilon}{{\varepsilon}}

\let\t\tensor
\let\p\partial
\def\dd{\mathrm{d}}
\def\ii{\mathrm{i}}
\def\ee{\mathrm{e}}
\newcommand{\vecb}[1]{\mathbf{#1}}
\newcommand{\deltaT}{\delta \mathrm{T}}

\let\t\tensor
\let\p\partial

\def\dd{\mathrm{d}}
\def\half{\tfrac{1}{2}}

\def\grav{\mathtt{g}} % gravitational acceleration
\def\poisson{\nu} % Poisson's ratio -- \nu and \sigma are both common
\newcommand{\shearmod}{\mu} % shear mudulus

\newcommand{\alphat}{\alpha}

\newcommand{\alpham}{\kappa}

\newcommand{\eszeta}{\zeta}

\def\permeability{\upmu}
\def\permittivity{\upepsilon}

\usepackage{siunitx}

\def\Pp{p}
\def\Qq{q}
\def\Kk{(u_{xx}+u_{yy})|_{r=0}}
\def\Bb{(u_{xx}-u_{yy})|_{r=0}}
\def\esGamma{\tilde \Gamma}

\usepackage{float, subcaption}

\usepackage{orcidlink}

\newcommand{\ptcnh}[1]{}% comments to be eventually read later

\begin{document}
\title{Elastically induced phase-shift and birefringence in optical fibers%
\thanks{Vienna preprint UWThPh 2025-2}
}

\author[1,2]{E.~Steininger \orcidlink{0000-0003-0092-3043}}
\author[1]{T.~Mieling \orcidlink{0000-0002-6905-0183}}
\author[1]{P.T.~Chru\'sciel \orcidlink{0000-0001-8362-7340}}
\affil[1]{University of Vienna, Faculty of Physics and Research Network TURIS, Boltzmanngasse~5, 1090 Vienna, Austria}
\affil[2]{University of Vienna, Faculty of Physics, Vienna Doctoral School in Physics, Boltzmanngasse 5, 1090 Vienna, Austria}

\date{\today}
\maketitle

\begin{abstract}
We compute how elastic deformations of optical fibers affect light propagation therein. 
Specifically, we consider differences in wave-guiding properties of straight fibers subject to different external temperatures, pressures, and gravitational fields.
This is done by solving, perturbatively to first order, the Maxwell equations in deformed and anisotropic fibers using a multiple-scales approximation scheme.  We derive explicit expressions for the induced phase shift and birefringence.
The phase shift can be expressed in terms of the average radial pressure, longitudinal tension, and change in temperature, while birefringence depends on the quadrupole of the external pressure distribution and the stresses on the axis of the fiber.
\end{abstract}
\tableofcontents

\section{Introduction}\label{sec:intro}

\begin{figure}[htb]
	\centering
	\includegraphics[width=0.5\columnwidth]{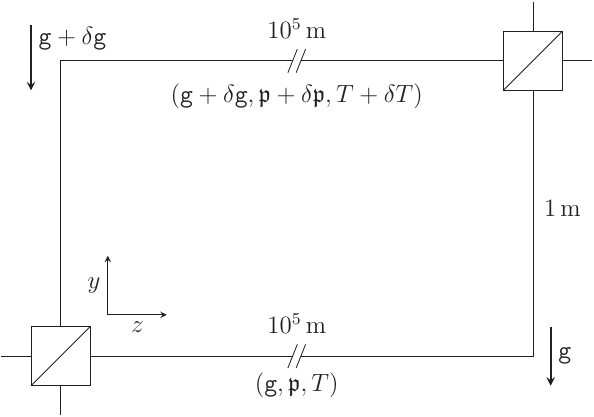}
	\caption{Schematic representation of the GRAVITES experiment \cite{Hilweg:2017ioz}. Light propagates through optical fibers in two vertically separated arms of an interferometer and traverses spools with a total fiber length of the order of $100\,\mathrm{km}$, accumulating different phases due to the gravitational redshift. We consider a simplified setup as in previous work \cite{BCS24}, where the spools are unwound and treated as straight fiber sections. We allow for a difference in gravitational acceleration $\delta \grav$, ambient pressure $\delta \mathfrak p$ and ambient temperature $\delta T$ at the spool locations, leading to elastic effects on the photon propagation. \copyright\  E.~Steininger, reproduced with permission. \cite{BCS24}}
	\label{fig:gravites_sketch}
\end{figure}

Long-baseline interferometry can be carried out in tabletop experiments using optical fibers, thus enabling the measurement of small perturbations in the dispersion relation of light (even at the single-photon level). Such phase shifts arising from the Sagnac effect have already been measured \cite{2019NJPh...21e3010F,2023PhRvR...5b2005C}, and an experiment is being built to measure such phase shifts arising from vertical displacements of one arm of the interferometer in the gravitational field of the Earth \cite{Hilweg:2017ioz,2024Polini}.
Compared to experiments involving light propagation in vacuum,  such experiments face the challenge of minimizing, controlling, and compensating for noise arising inside the fiber, or arising from the response of fibers to changes, drifts, and fluctuations of ambient conditions \cite{2022Optic...9.1238H}.
In particular, for experiments that modulate their signal by moving one or both interferometer arms (as, e.g., in Refs.~\cite{Hilweg:2017ioz,2024Polini}), local variations in pressure, temperature, and local gravity lead to elastic deformations of the fibers that influence light propagation through photoelastic effects.
These effects need to be modeled accurately in order to extract the effect which one aims to measure. The aim of this work is to carry this out, with focus on the effects relevant for the planned GRAVITES experiment \cite{Hilweg:2017ioz}.

A simplified schematic diagram of this last experiment is shown in Fig.~\ref{fig:gravites_sketch}.
The interferometer aims to measure phase shifts induced by Earth's gravity on entangled multi-photon states propagating in optical fibers. To modulate this effect, one interferometer arm is gradually lifted by $1\,\mathrm{m}$ relative to a stationary reference arm. If the fiber length is held constant in the process, the gravitational redshift implies a difference in phase for the two interferometer arms, which results in a height-dependent change in detection statistics of individual photons at the two output ports of the final beam-splitter.
Now, the gravitational field in the laboratory changes with height. Further, in general, the pressures and temperatures at the two interferometer arms are not  equal. The differences in gravitational acceleration, ambient pressure, and ambient temperature, will be denoted by $\delta\grav$, $\delta\mathfrak p$, and $\delta T$.
We wish to determine  here the changes of the Maxwell field in a cylindrical waveguide as functions of these variables.

As such, the elastic response of (straight) optical fibers to variations in gravity, pressure, and temperature, was accurately modeled in a previous work \cite{BCS24}.
We build upon these results by solving Maxwell’s equations in so-deformed fibers to determine the effect of variations of ambient parameters on light propagation in single-mode fibers. This is done using the perturbative scheme developed in Refs.~\cite{2023Mieling,2023PhRvR...5b3140M,2024arXiv241023048M}.

This paper is organized as follows.
In Sect.~\ref{sec:pert_fiber_optics} we review the formulation of Maxwell’s equations in linear dielectrics in terms of gauge-fixed wave equations for the electromagnetic potential.
These equations can be solved explicitly for the case of undeformed and unstrained step-index fibers, yielding the standard mode solutions. Following the approach of Refs.~\cite{2024arXiv241023048M,2023Mieling} and restricting to  single-mode fibers, we extend the calculations there to include generic perturbation terms, and derive the general solution to first order in perturbation theory.
This   result is then used in the following sections to describe the phase shift and birefringence induced in the fiber by elastic  deformations  of any origin.

In Section~\ref{sec:elasticity} we review the  elastic deformations of optical fibers derived in Ref.~\cite{BCS24}, and derive the induced correction terms in the electromagnetic field equations describing light propagation in such fibers.
These corrections arise directly from the geometric fiber deformation (that alters the shape of the core-cladding interface), and indirectly through the induced anisotropy (arising from the strain due to photoelasticity). To first order in perturbation theory  these effects add and can thus be computed separately: In Section~\ref{sec:displacement} we describe the purely geometric deformation effects, and in Section~\ref{sect:photoelasticity} we discuss the photoelastic effects.
The resulting expressions for the phase shifts and birefringence contain integrals that need to be evaluated numerically.

\begin{figure}[t]
	\centering
	\begin{subfigure}[t]{.45\columnwidth}
		\includegraphics[width=\columnwidth]{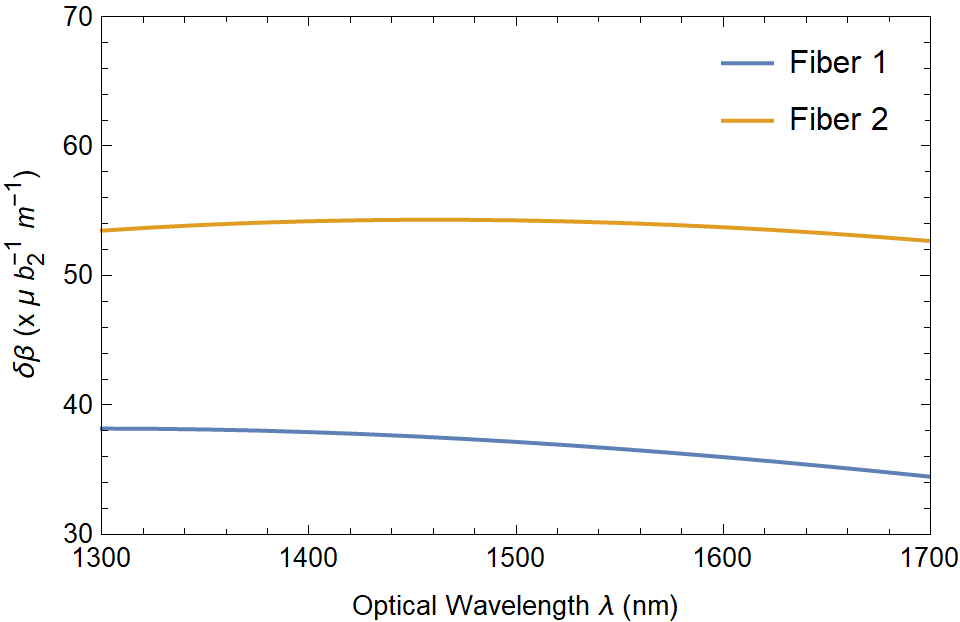}
		\caption{}
		\label{fig:birefr:deformation:wavelength}
	\end{subfigure}
	\hfill
	\begin{subfigure}[t]{.45\columnwidth}
		\includegraphics[width=\columnwidth]{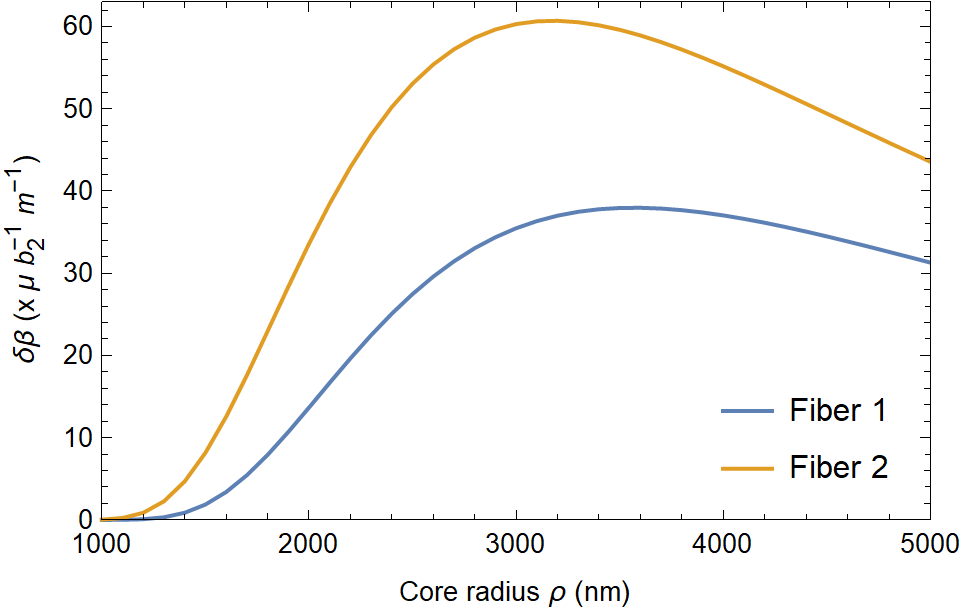}
		\caption{}
		\label{fig:birefr:deformation:radius}
	\end{subfigure}
	\\
	\begin{subfigure}[t]{.45\columnwidth}
		\includegraphics[width=\columnwidth]{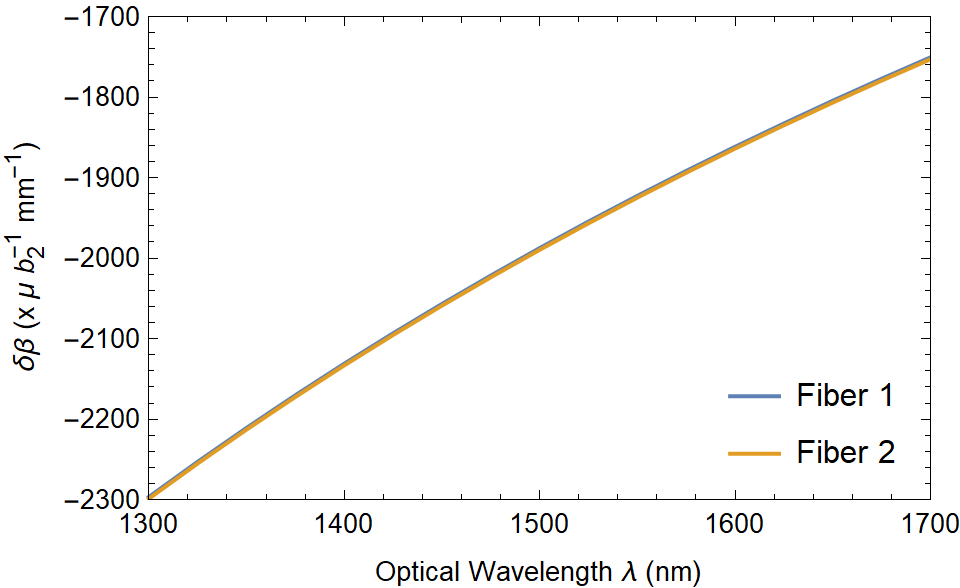}
		\caption{}
		\label{fig:birefr:photoelastic:wavelength}
	\end{subfigure}
	\hfill
	\begin{subfigure}[t]{.45\columnwidth}
		\includegraphics[width=\columnwidth]{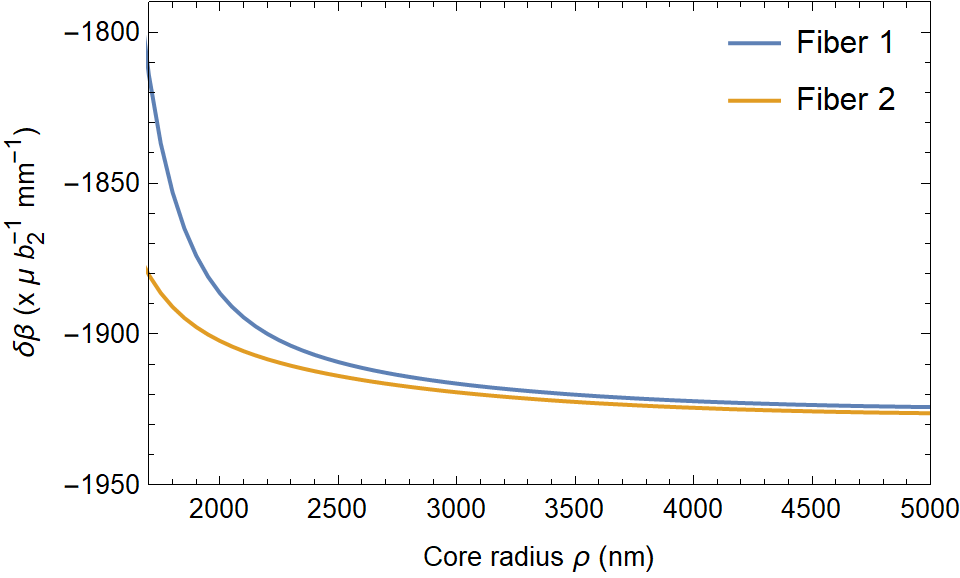}
		\caption{}
		\label{fig:birefr:photoelastic:radius}
	\end{subfigure}
	\caption{
		Varying the wavelength (left column) or the fiber core radius (right column) changes the   birefringence  induced by the elastically deformed core-cladding interface (\ref{fig:birefr:deformation:wavelength} and \ref{fig:birefr:deformation:radius}) and the birefringence caused by the photoelastic material response (\ref{fig:birefr:photoelastic:wavelength} and \ref{fig:birefr:photoelastic:radius}).
		Fiber 1 describes a fiber with parameters as listed in Tables~\ref{tab:SiO2}--\ref{tab:parGRAVITES}, and Fiber 2 has refractive indices $n_1 = 1.4715$ and $n_2 = 1.4648$ instead, with the fibers identical otherwise.
		The parameter $b_2$ quantifies the strain in the waveguide as defined in \eqref{14I15.11}.
		In Figure~\ref{fig:birefr:photoelastic:wavelength}, and in the physically relevant region of Figure~\ref{fig:birefr:photoelastic:radius}, both fiber’s responses are identical for all practical purposes.
	}
	\label{fig:birefringence}
\end{figure}

\begin{figure}[ht]
	\centering
	\begin{subfigure}[t]{.466\columnwidth}
		\includegraphics[width=\columnwidth]{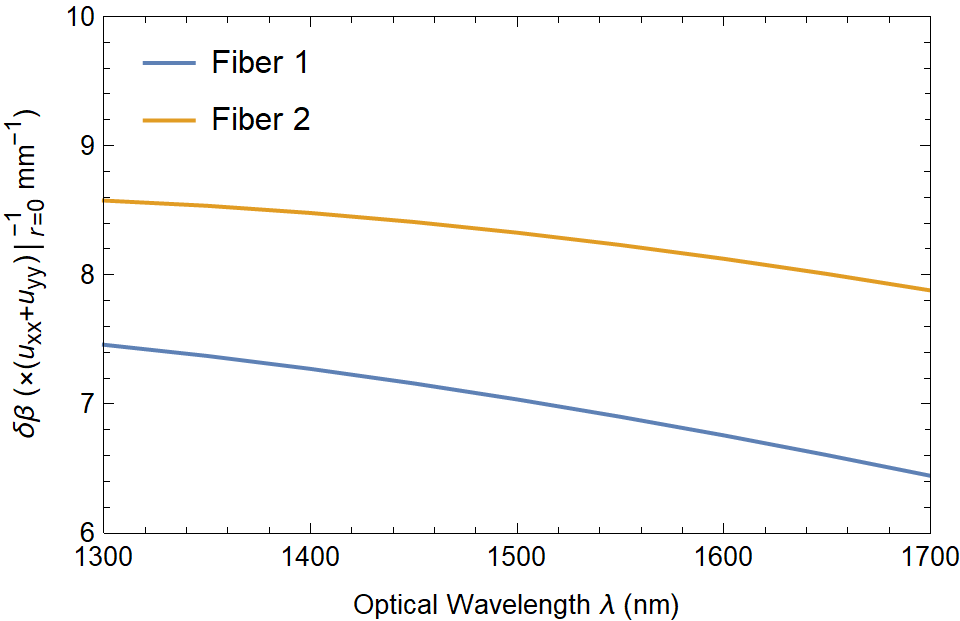}
		\caption{}
		\label{fig:phase:displacement:wavelength}
	\end{subfigure}
	\hfill
	\begin{subfigure}[t]{.45\columnwidth}
		\includegraphics[width=\columnwidth]{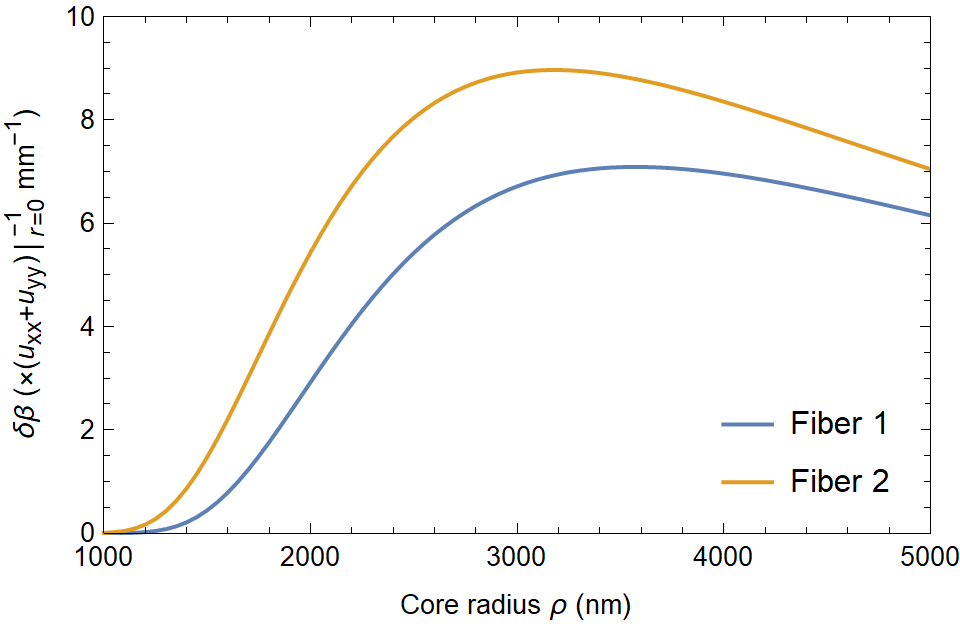}
		\caption{}
		\label{fig:phase:displacement:radius}
	\end{subfigure}
	\begin{subfigure}[t]{.466\columnwidth}
		\includegraphics[width=\columnwidth]{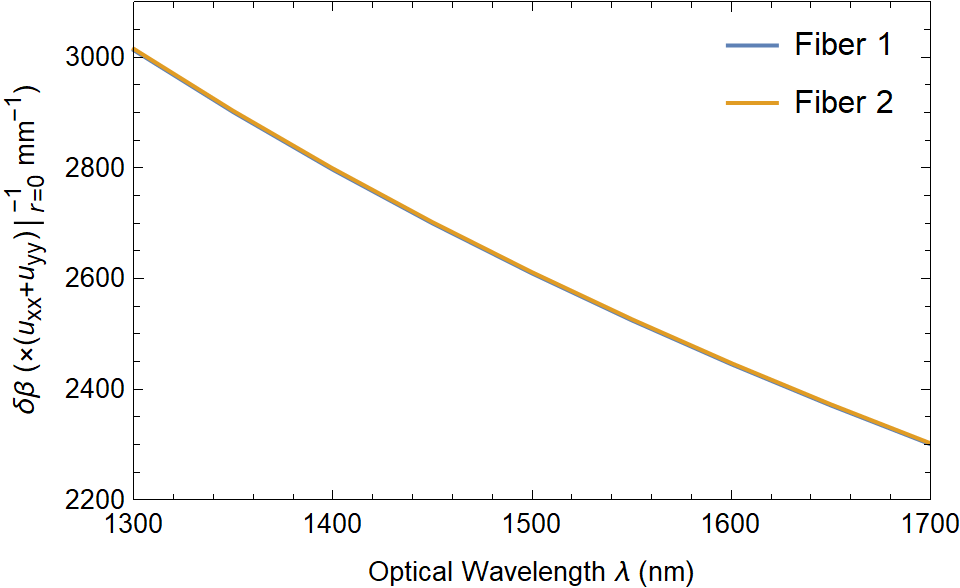}
		\caption{}
		\label{fig:phase:photoelastic:wavelength}
	\end{subfigure}
	\hfill
	\begin{subfigure}[t]{.45\columnwidth}
		\includegraphics[width=\columnwidth]{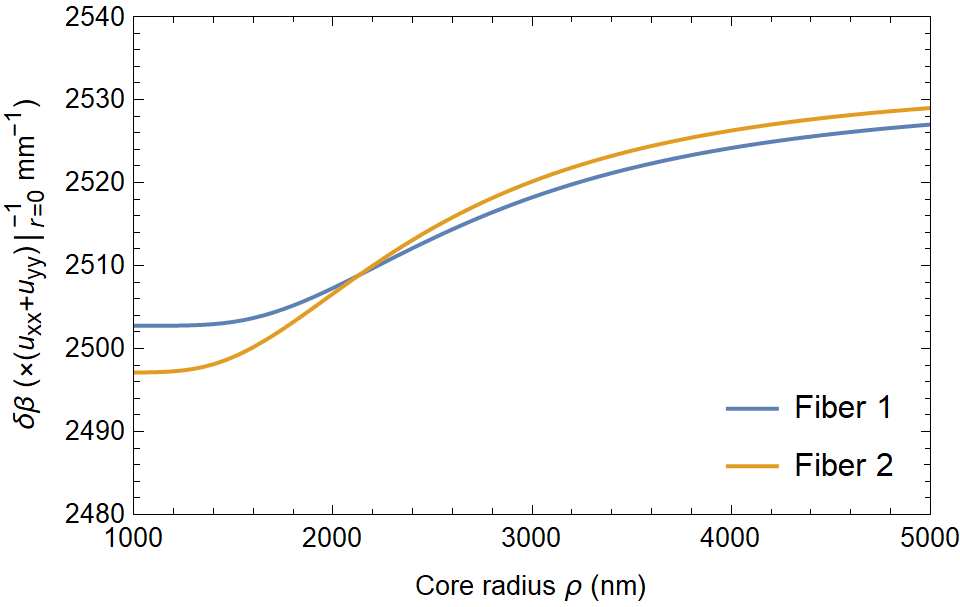}
		\caption{}
		\label{fig:phase:photoelastic:radius}
	\end{subfigure}
	\caption{
		Figures~\ref{fig:phase:displacement:wavelength} and \ref{fig:phase:displacement:radius} show the dependence of the phase shift on the elastic deformation of the core-cladding interface as functions of the optical wavelength and the fiber’s core radius.
		Figures~\ref{fig:phase:photoelastic:wavelength} and \ref{fig:phase:photoelastic:radius}, on the other hand, show the dependence of the photoelastically induced phase shift on these parameters.
		The material parameters are the same as in Figure~\ref{fig:birefringence}.
		 In Figure~\ref{fig:phase:photoelastic:wavelength}, and in the physically relevant region of Figure~\ref{fig:phase:photoelastic:radius}, the responses of both fibers are identical for all practical purposes. 
	}
	\label{fig:phase shifts}
\end{figure}

Figures \ref{fig:birefringence} and \ref{fig:phase shifts} summarize the results of our analysis and can be viewed as the main result of this paper.

In Section~\ref{sec:numerics} we provide numerical results with parameters relevant for the GRAVITES experiment.


Our calculations  are presented in units in which the vacuum permeability $\upmu_0$,  the vacuum permittivity $\upepsilon_0$, and hence also the speed of light in vacuum $c$, are set to unity.
Note that the final formulæ do not involve these constants.
Coordinate indices are denoted by $\mu$, $\nu$, $\ldots$, and range from $0$ to $3$, while the frame indices are denoted by $a$, $b$, $\ldots$, and range over the set $\{t, z, \sharp, \flat\}$, compare Eq.~\eqref{eq:complex frame} below.
Implicit summation over repeated indices is employed throughout.

\section{Perturbative fiber optics}\label{sec:pert_fiber_optics}

Let us start by introducing the formalism used to describe electromagnetic modes in deformed single-mode fibers.
To solve Maxwell’s equations in such fibers  it is useful to formulate the equations in curvilinear coordinates adapted to the deformation of the fiber.
For this in Section~\ref{s:maxwell general theory} we recall a generally covariant formulation of Maxwell’s equations in linear dielectrics.
As these equations are generally not solvable explicitly, perturbation techniques are required.
In our case, the unperturbed reference problem is that of an undeformed fiber, the modes of which are described in Section~\ref{sec:unperturbed}.
Based on this, in Section~\ref{sec:multiplescale} we present  an approximation scheme that allows to compute perturbations of such modes.
This scheme is general in that it applies to arbitrary small deviations from unperturbed field equations, regardless of their origin, as long as the perturbations are invariant under translations along the fiber.
The results in this section will be used in Section~\ref{sec:elasticity} to determine the effects of internal deformations and of stress-induced anisotropy.

\subsection{Maxwell’s equations in linear media}
\label{s:maxwell general theory}

Maxwell’s equations without sources can be written in generally covariant form as
\begin{subequations}
	\begin{align}
		\label{eq:maxwell:homogeneous}
		\t\nabla{_[_\mu} \t F{_\mu_\nu_]} &= 0\,,
		\\
		\label{eq:maxwell:inhomogeneous}
		\t\nabla{_\mu} \t G{^\mu^\nu} &= 0\,,
	\end{align}
\end{subequations}
where $\t F{_\mu_\nu}$ and $\t G{^\mu^\nu}$ are antisymmetric tensors describing the electromagnetic field strength and excitation, where $\t\nabla{_\mu}$ is the covariant derivative associated to the space-time metric tensor $\t g{_\mu_\nu}$, and where square brackets indicate antisymmetrization (cf.\ e.g.~\cite{Post_FormalStructure}).
To obtain a closed system it is necessary to formulate a \emph{constitutive law} that relates $\t G{^\mu^\nu}$ and $\t F{_\mu_\nu}$.
For linear and non-dispersive media, the standard form of this law takes the form
\begin{align}\label{18V24.1}
	\t G{^\mu^\nu}
		&= \t \chi{^\mu^\nu^\rho^\sigma} \t F{_\rho_\sigma}\,,
\end{align}
with the \emph{constitutive tensor} $\chi^{\mu\nu\rho\sigma}$ exhibiting the symmetries
\begin{align}
	\t \chi{^\mu^\nu^\rho^\sigma}
		= + \t \chi{^\rho^\sigma^\mu^\nu}
		= - \t \chi{^\nu^\mu^\rho^\sigma}
		= - \t \chi{^\mu^\nu^\sigma^\rho}\,.
\end{align}
For isotropic dielectrics with four-velocity $\t u{^\mu}$, permittivity $\permittivity$, and permeability $\permeability$, it can be shown \cite{Gordon23} that the constitutive tensor takes the form
\begin{align}
	\t*\chi{_{\text{isotropic}}^\mu^\nu^\rho^\sigma}
		&= \tfrac{1}{2 \permeability} (\t\gamma{^\mu^\rho} \t\gamma{^\nu^\sigma} - \t\gamma{^\mu^\sigma} \t\gamma{^\nu^\rho})\,,
\end{align}
where $\t\gamma{^\mu^\nu}$ is \emph{Gordon’s optical metric}
\begin{align}
	\t\gamma{^\mu^\nu}
		&= \t g{^\mu^\nu} + (1- n^2) \t u{^\mu} \t u{^\nu}\,,
\end{align}
in which $n = \sqrt{\permittivity \permeability}$ is the refractive index.

Optical fibers are often modeled as linear isotropic dielectrics with $\permeability = 1$ \cite{2005Liu}.
However, to describe photoelastic effects in fibers, we will allow more general constitutive tensors $\t*\chi{^\mu^\nu^\rho^\sigma}$ that reduce to $\t*\chi{_{\text{isotropic}}^\mu^\nu^\rho^\sigma}$ with $\permeability = 1$ in the absence of deformations and stresses.

The following calculations are carried out in terms of an electromagnetic potential $\t A{_\mu}$ such that 
\begin{equation}
 \t F{_\mu_\nu} = \t\nabla{_\mu} \t A{_\nu} - \t\nabla{_\nu} \t A{_\mu}
  \,.
   \label{13I25.4}
\end{equation}
In this formulation Eq.~\eqref{eq:maxwell:homogeneous} is identically satisfied, and Eq.~\eqref{eq:maxwell:inhomogeneous} does not directly provide  well-posed evolution equations due to the gauge redundancy $\t A{_\mu} \to \t A{_\mu} + \t\partial{_\mu} \lambda$.
Following Refs.~\cite{2022PhRvA.106f3511M,2023Mieling,2024arXiv241023048M} we therefore consider the gauge-fixed Lagrangian
\begin{align}
	\mathcal L
		&= - \quarter \t\chi{^\mu^\nu^\rho^\sigma} \t F{_\mu_\nu} \t F{_\rho_\sigma}
		- \half \t*\chi{_{\text{gauge}}^2}\,,
	&
	&\text{with}
	&
	\t*\chi{_{\text{gauge}}}
		&= \t\nabla{_\mu}(\t\gamma{^\mu^\nu} \t A{_\nu})\,,
\end{align}
where $\t\gamma{^\mu^\nu}$ is the Gordon metric with $\permeability = 1$ and $\permittivity = n^2$ being the permittivity of the undeformed  reference configuration, without deformations and stresses.
The corresponding Euler–Lagrange equations
\begin{align}
	\label{eq:maxwell + gauge}
	\t\nabla{_\mu} \t G{^\mu^\nu}
		+ \t\gamma{^\mu^\nu} \t\nabla{_\mu} \t*\chi{_{\text{gauge}}}
		&= 0
\end{align}
reduce to Eq.~\eqref{eq:maxwell:inhomogeneous} whenever $\t*\chi{_{\text{gauge}}}$ vanishes, and are well-posed irrespective of this gauge condition \cite{2023Mieling}.

In everything that follows the metric is taken to be the Minkowski metric, which we denote by $\eta_{\mu\nu}$, and the four-velocity vector is constant throughout the medium.
Hence, in the case of an idealized step-index fiber, where $\t\chi{^\mu^\nu^\rho^\sigma} = \t*\chi{^\mu^\nu^\rho^\sigma_{\text{isotropic}}}$ with $\permeability = 1$ and $\permittivity = n^2$ being locally constant ($n$ takes different values in the fiber’s core and cladding),  in Minkowski spacetime and in inertial coordinates Eq.~\eqref{eq:maxwell + gauge} reduces to
\begin{align}
	\label{10VI24.1}
	(- n^2 \partial_t^2 + \partial_x^2 + \partial_y^2 + \partial_z^2) \t A{_\mu} = 0\,.
\end{align}

\subsection{Solutions for ideal optical fibers}\label{sec:unperturbed}

In this section we review  the mode solutions to Eq.~\eqref{10VI24.1} for straight step-index fibers with circular cross-sections. These serve as unperturbed reference solutions to the perturbative scheme described in Section~\ref{sec:multiplescale}. Whereas the description of such modes at the level of the field strength is described in textbooks \cite{2005Liu}, the following calculation is carried out using the electromagnetic potential following Ref.~\cite{2022PhRvA.106f3511M}.
To distinguish the reference solution derived here from the actual solution for deformed fibers, the solutions to the unperturbed problem will be marked by a ring diacritic ($\mathring{\_}$).

The unperturbed optical fiber is modelled as a cylindrical dielectric fiber with a core  of radius $\rho$ and refractive index $n_1$ that is surrounded by an annular dielectric cladding of outer radius $a$ and refractive index $n_2 < n_1$.
While $a$ is finite for the purpose of the elasticity calculations, because of the exponential decay of the electromagnetic field in the cladding we use the approximation $a=\infty$ when solving Maxwell’s equations.

An efficient approach to this problem proceeds by solving 
Eq.~\eqref{10VI24.1} in regions of constant refractive index independently and applying matching conditions at the interface.
Due to the symmetry of the problem, the   calculations to follow are carried out using cylindrical coordinates $(t, r, \theta , z)$.
Following Ref.~\cite{2022PhRvA.106f3511M}, mode solutions are obtained by decomposing the potential $\t A{_\mu}$ using the complex frame
\begin{subequations}
\label{eq:complex frame}
\begin{align}
	e_t
		&= \p_t\,,
	&
	e_\sharp
		&= \tfrac{1}{\sqrt 2} (\p_r + \tfrac{\ii}{r}\p_\theta)\,,
	\\
	e_z
		&= \p_z\,,
	&
	e_\flat
		&= \tfrac{1}{\sqrt 2} (\p_r - \tfrac{\ii}{r}\p_\theta)\,,
\end{align}
\end{subequations}
and using the ansatz
\begin{equation}\label{2XI24.1}
	\mathring A_b = \mathring a_b (r) \ee^{\ii (\beta z + m \theta - \omega t)}\,,
\end{equation}
so that the wave equations for the frame components $\mathring a_b$ decouple.
Here, $\omega$ is the angular frequency of the mode, $\beta$ its propagation constant, and $m$ is the azimuthal mode index.
Writing
\begin{align}
	\mathring a 
		&= (\mathring a_b)
		\equiv (\mathring a_t,\quad \mathring a_z,\quad \mathring a_\sharp,\quad \mathring a_\flat)\,,
\end{align}
the system~\eqref{10VI24.1} is equivalent to
\begin{equation}\label{10II24.1}
	\mathscr{H}_m \mathring a 
		\coloneqq
 \big(H_m \mathring a_t, \quad H_m \mathring a_z, \quad H_{m+1} \mathring a_\sharp, \quad H_{m-1}\mathring a_\flat
 \big)
		= 0\,,
\end{equation}
where $H_m$ is the Helmholtz operator
\begin{equation}
	H_m
		= \frac{\p^2}{\p r^2}
		+ \frac{1}{r} \frac{\p}{\p r}
		- \frac{m^2}{r^2}
		+ n^2 \omega^2 - \beta^2\,.
\end{equation}
The requirements of regularity at the origin $r=0$ and of exponential fall-off as $r\rightarrow \infty$ imply that solutions of \eqref{10II24.1} can be expressed in terms of the functions
\begin{equation}\label{eq:15II24.3}
	f_m (c_1 , c_2 , r) \coloneqq \Bigg\{ \begin{array}{lr} c_1 J_m (U r) & r<\rho\,, \\ c_2 K_m (W r) & r>\rho\,, \end{array}
\end{equation}
where $c_1$ and $c_2$ are arbitrary constants, $J_m$ are Bessel functions of first kind, $K_m$ are modified Bessel functions of second kind, and the parameters $U$ and $W$ are defined as
\begin{align}
	\label{eq:UW}
	U
		&= \rho \sqrt{ n_1^2 \omega^2 - \beta^2 }\,,
	&
	W
		&= \rho \sqrt{ \beta^2 - n_2^2 \omega^2 }\,.
\end{align}
Using this notation, the relevant solutions to Eq.~\eqref{10II24.1} take the form
\begin{equation}
	\mathring a
		= \mathscr{F}_m (\mathring{\vecb c})
		\coloneqq
		\big(
			f_m(c_1 , c_5, r),\quad
			f_m (c_2 , c_6, r),\quad
			f_{m+1}(c_3 , c_7, r),\quad
			f_{m-1} (c_4, c_8, r)
		\big)
 \,,
\label{eq:15IV24.1}
\end{equation}
with 
\begin{align}
	\mathring {\vecb c} = (c_1, \dots , c_8)
\end{align}
denoting a set of constants that needs to be determined from the matching conditions at the core-cladding interface located at $r = \rho$.
Specifically, we require continuity of the following field components, as discussed in Section II.D.\ of Ref.~\cite{2022PhRvA.106f3511M}
\begin{align}\label{eq:junction_conditions}
	\llbracket \chi_\text{gauge} \rrbracket &= 0\,,
	&
	\llbracket \t A{_\mu} \rrbracket &= 0\,,
	&
	\llbracket \t G{^\mu^\nu} \rrbracket \t n{_\nu} &= 0\,,
\end{align}
in which $\llbracket f\rrbracket  = (f_\text{core} - f_\text{cladding}) |_{r=\rho}$ denotes the jump of a function at the interface and $\t n{_\mu}$ denotes any normal to that interface (due to linearity neither the sign nor the normalization are relevant).
For the ansatz \eqref{2XI24.1} these conditions can be written concisely as $\Gamma_m [\mathring a] = 0$, where $\Gamma_m [\mathring a]$ is the eight-component vector
\begin{eqnarray}\label{24VI24.1}
	\Gamma_m [\mathring a] =
	\begin{pmatrix}
	 	\llbracket \mathring a_0 \rrbracket \\
		\llbracket \mathring a_\sharp \rrbracket \\
	 	\llbracket \mathring a_\flat \rrbracket \\
	  	\llbracket \mathring a_z \rrbracket \\
		\llbracket n^2 (\p_r \mathring a_0 + \frac{\ii \omega}{\sqrt 2} (\mathring a_\sharp + \mathring a_\flat))\rrbracket \\
		\llbracket \p_r (\mathring a_\sharp - \mathring a_\flat) + \frac{m+1}{ r} \mathring a_\sharp + \frac{m-1} {r} \mathring a_\flat  \rrbracket\\
		\llbracket \p_r \mathring a_z \rrbracket \\
		\llbracket \ii \omega n^2 \mathring a_0 + \frac{1}{\sqrt 2} \left(\p_r (\mathring a_\sharp + \mathring a_\flat) +\frac{m+1}{ r}\mathring a_\sharp  -  \frac{m-1} {r}\mathring a_\flat \right) \rrbracket\\
	\end{pmatrix}\,,
\end{eqnarray}
where we kept some of the continuous terms as this simplifies the algebra later.

For solutions of the form \eqref{eq:15IV24.1}, the interface conditions can be written in terms of matrices acting on the vector $\mathring{\vecb c}$:
\begin{equation}\label{10VI24.2}
	 \Gamma_m [\mathring a] = M_m N_m \mathring{\vecb c}\,,
\end{equation}
where
\begin{equation}
	\label{eq:matrix M}
	 M_m =
	 \begin{psmallmatrix}
			 1 & 0 & 0 & 0 & -1 & 0 & 0 & 0 \\
 			0 & 1 & 0 & 0 & 0 & -1 & 0 & 0 \\
 			0 & 0 & \frac{m}{U}-\mathcal{J} U & 0 & 0 & 0 & \mathcal{K} W-\frac{m}{W} & 0 \\
			 0 & 0 & 0 & \frac{m}{U}+\mathcal{J} U & 0 & 0 & 0 & \frac{m}{W}+\mathcal{K} W \\
 			\mathcal{J} n_1^2 U^2 & 0 & \frac{\ii n_1^2 \omega  (m-\mathcal{J} U^2)}{\sqrt{2} U} & \frac{\ii n_1^2 \omega  (m+\mathcal{J}U^2)}{\sqrt{2} U} & -\mathcal{K} n_2^2 W^2 & 0 &
   \frac{\ii n_2^2 \omega  (\mathcal{K} W^2-m)}{\sqrt{2} W} &
   \frac{\ii n_2^2 \omega  (m+\mathcal{K} W^2)}{\sqrt{2} W} \\
 0 & 0 & U & U & 0 & 0 & W & -W \\
 0 & \mathcal{J} U^2 & 0 & 0 & 0 & -\mathcal{K} W^2 & 0 & 0 \\
 \ii n_1^2 \omega  & 0 & \frac{U}{\sqrt{2}} & -\frac{U}{\sqrt{2}} & - \ii n_2^2
   \omega  & 0 & \frac{W}{\sqrt{2}} & \frac{W}{\sqrt{2}} \\
\end{psmallmatrix}\,,
\end{equation}
with
\begin{align}
	\mathcal J &= \frac{J_m'}{U J_m} \,,
	&
	\mathcal K &= \frac{K_m'}{W K_m}\,,
\end{align}
and
\begin{equation}
	\label{eq:matrix N}
	N_m = \operatorname{diag} (J_m \quad J_m \quad  J_m \quad  J_m \quad  K_m \quad K_m \quad K_m\quad K_m )\,.
\end{equation}
For conciseness, we suppress here the arguments of the Bessel functions, which are evaluated at the interface and read $J_m (U \rho)$ and $K_m (W \rho)$, respectively.

\begin{figure}
  \centering
  \includegraphics[width=.7\textwidth]{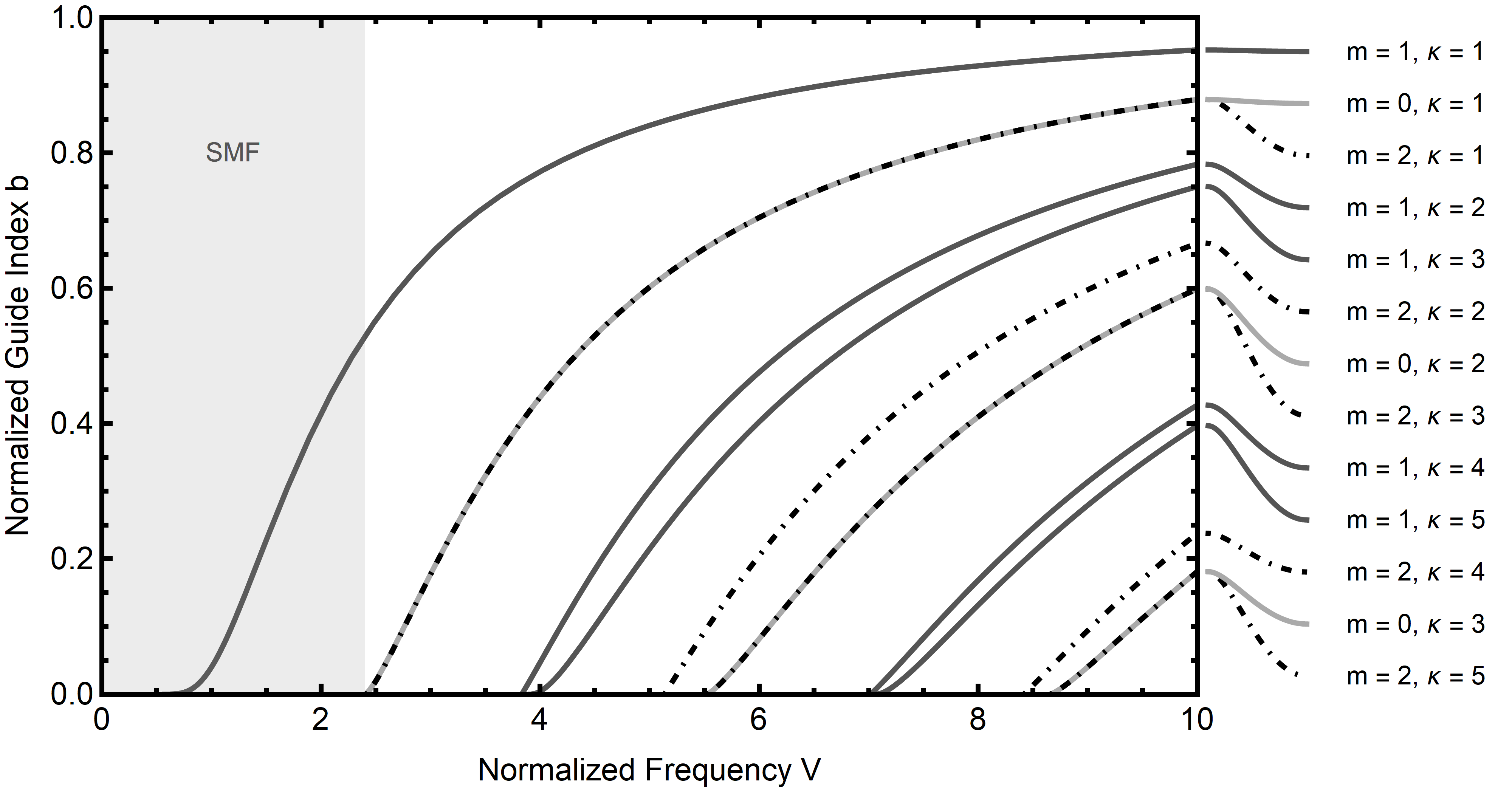}
  \caption{Mode diagram for an optical fiber with refractive indices $n_1 = 1.4712$ and $n_2 = 1.4659$. The parameter $\kappa$ enumerates distinct roots for a fixed azimuthal mode number $m$. Note that in the shaded region only a single propagating mode exists for a given normalized frequency. This region corresponds to single mode optical fibers (SMF).}
  \label{fig:unperturbed_modes}
\end{figure}

Non-trivial solutions $\mathring{\vecb c}$ to $\Gamma_m [\mathring a] = 0$ exist if and only if $\det M = 0$. One can show that this determinant factorizes into a term whose roots yield physical solutions and a term whose roots correspond to gauge and ghost solutions that have either vanishing field strength $F_{\mu\nu}$ or violate the gauge condition $\t*\chi{_{\text{gauge}}} = 0$ \cite{2022PhRvA.106f3511M}.
The equation leading to physical solutions reads
\begin{equation}
	(\mathcal J + \mathcal K) ( n_1^2 \mathcal J + n_2^2 \mathcal K) = m^2 \frac{ ( U^2 + W^2 )^2 }{ U^4 W^4 } \frac{ \beta^2 }{ \omega^2 } \,,
\label{15II24.4}
\end{equation}
where individual roots of this transcendental equation yield propagating modes.
Defining the normalized frequency
\begin{equation}
	V = \omega \rho \sqrt{n_1^2 - n_2^2}\,,
\end{equation}
and the normalized guide index $b$ implicitly by 
\begin{align}
	U
		&= \sqrt{1-b} V\,,
	&
	W
		&= \sqrt{b} V\,,
\end{align}
The numerical solution to Eq.~\eqref{15II24.4} with $n_1 = 1.4712$ and $n_2 = 1.4659$ is presented in Figure~\ref{fig:unperturbed_modes}, compare Ref.~\cite{2005Liu}.
Note in particular the shaded area in this diagram, where, for a given normalized frequency, there exist only roots for \eqref{15II24.4} with mode numbers $m = \pm 1$ (the curves are indifferent to the sign of $m$).
Typical applications of single-mode fibers take place in this regime.

For a given simple root of the determinant, the kernel spans a one-dimensional subspace, leading to solutions $\mathring{\vecb c}$ being determined up to an overall constant. This becomes important when one tries to define the polarization state of a superposition of simultaneously propagating solutions, 
as the solution vectors spanning the $m=\pm1$-subspaces can be scaled independently \cite{2024arXiv241023048M}.
In the case of single-mode fibers the $m=\pm1$ solutions share the same root, and inspection of \eqref{eq:matrix M} shows that a change of sign for the mode number $m$ corresponds to some sign changes and re-ordering of the entries, namely
\begin{equation}
	\mathring{\vecb c}^- = \begin{pmatrix} -\mathring c^+_1 & \mathring c^+_2 & \mathring c^+_4 & \mathring c^+_3 & \mathring c^+_5 & \mathring c^+_6 & \mathring c^+_8 & \mathring c^+_7 \end{pmatrix}\,,
\end{equation}
up to a multiplicative factor that we chose to be equal to one. 

\subsection{Multiple-scales analysis}
\label{sec:multiplescale}

Basing on the above description of light propagation in ideal optical fibers, in this section we derive a general formula that describes the polarization dynamics in deformed optical fibers that are invariant under translations along the fiber.
This result is applied in Section~\ref{sec:elasticity} to compute elastically induced phase and polarization dynamics resulting from stress-induced optical inhomogeneity and anisotropy, as well as from deformations of the core-cladding interface.
Denoting by $\varepsilon \ll 1$ a dimensionless expansion parameter associated to such perturbations, corrections to the field equations \eqref{eq:maxwell + gauge} can generally be written as
\begin{equation}
	\label{eq:perturbed wave equation}
	\Box_\gamma A_\nu + \varepsilon \Sigma A_\nu = 0\,,
\end{equation}
where $\Sigma$ is an operator that describes all first-order perturbations of the wave equation that arise from elasticity, and we neglect second order $\varepsilon$-terms throughout.
For our purposes there is no loss of generality in working with coordinates such that the core-cladding interface lies at $r = \rho$, as perturbed fiber geometries can be transformed to this setting by a change of coordinates (see Section~\ref{sec:displacement} for details).
This method facilitates the implementation of junction conditions at the core-cladding interface.

Restricting to the single-mode fiber solutions of Section~\ref{sec:unperturbed}, the $O(\varepsilon^0)$-modes
 have azimuthal mode number $m=\pm 1$.
 The remaining modes are a priori $O(\varepsilon)$.
 As shown in Section III.C.\ of Ref.~\cite{2024arXiv241023048M}, the propagation of these leading modes in \eqref{eq:perturbed wave equation} is governed by those $\Sigma$-terms that are isotropic (i.e., independent of $\theta$) and those having a relative azimuthal mode number of $\Delta m = \pm 2$. 
 Higher-order perturbation theory would allow a wider range of relative azimuthal mode numbers, since multiple combinations can again result in contributions to these leading order modes, see Refs.~\cite{2023PhRvR...5b3140M,2024arXiv241023048M} for such calculations.

Consider the unperturbed single-mode solution consisting of a superposition of $m = \pm1$ modes of the form \eqref{2XI24.1}:
\begin{equation}
	\mathring A=  \mathcal{A}\sum_{m=\pm 1} \mathscr{J}_m \mathscr{F}_m (\mathring{ \vecb c}_m)  \ee^{\ii (\beta z + m \theta - \omega t)}
 \,,
\end{equation}
where we factored-out an overall amplitude $\mathcal A$, so that the coefficients $\mathscr J_\pm$ are normalized to $|\mathscr J_+|^2 + |\mathscr J_-|^2 = 1$.
As shown in Ref.~\cite{2022PhRvA.106f3511M}, the coefficients $\mathscr J_\pm$ can be interpreted as projections of a complex Jones vector $\mathscr J = \mathscr J_x e_x + \mathscr J_y e_y$, describing the polarization of the solution, onto the complex basis $e_\pm = \tfrac{1}{\sqrt 2}(e_x \mp \ii e_y)$.
This translates into $\mathscr J_\pm = \tfrac{1}{\sqrt 2}(\mathscr J_x \mp \ii \mathscr J_y)$, and hence 
\begin{align}
	\mathscr J
		\equiv \mathscr J_x e_x + \mathscr J_y e_y  
		\equiv \mathscr J_+ e_- + \mathscr J_- e_+ 
	\,,	
\end{align}
so that $\mathscr J_+$ and $\mathscr J_-$ describe, respectively, the contributions of right-handed and left-handed circular polarizations. While the fields $\mathcal A$ and  $\mathscr{J}_m$ are constant in the unperturbed case, they satisfy propagation equations along the fiber in the perturbed case, and in fact the main aim of this work is to derive these equations in the current setting.

Following Ref.~\cite{2023PhRvR...5b3140M}  we describe gradual changes in the electromagnetic field along the fiber using the multiple-scales method. This proceeds by introducing an additional length variable $\eszeta = \varepsilon z$.
Using $z$ and $\eszeta$ simultaneously allows separating short-distance dynamics, involving distances of the order of the optical wavelength, from that on larger length scales which are determined by the deformations of the fiber.
We thus consider the following ansatz for the perturbed wave equation
\begin{equation}
	A_b
		= \sum_{m=\pm 1} \left( \mathring a_b^{m} (r,\eszeta)  + \varepsilon \tilde a_b^{m} (r,\eszeta) + O(\varepsilon^2)\right) \ee^{\ii (\beta z + m \theta - \omega t)}\,,
\end{equation}
where tilde diacritics ($\tilde \_$) denote first-order perturbation terms.
Note that the expansion above is to be understood in terms of the multiple-scales method, and \emph{not} as a Taylor series. We will thus \emph{not} approximate functions of $\eszeta$ by $f(\eszeta) = f(0) + \varepsilon z f'(0) + O(\varepsilon^2)$, see Chapter 11 of \cite{1978BenderOrszag} for a textbook exposition.

The wave equation splits then into terms proportional to $  \ee^{\ii (\pm 1 \pm \Delta m)\theta}$, where $\Delta m$ is the relative azimuthal mode number induced by the perturbations. 
To describe the evolution of the polarization along the fiber it suffices to consider the terms involving $  \ee^{\pm\ii\theta}$, leading to a coupled set of two equations
\begin{subequations}
\label{eq:15IV24.2}
\begin{align}
	\mathscr{H}_+ \mathring a_b^{+} + \varepsilon \left( \mathscr{H}_+ \tilde a_b^{+} + 2 \ii \beta \p_\eszeta \mathring a_b^{+} + \Sigma_b^{(+,0)}(\mathring a^{+})  +  \Sigma_b^{(+,2)}(\mathring a^{-})\right) &= 0\,,
\label{eq:15IV24.2a}
	\\
	\mathscr{H}_- \mathring a_b^{-} + \varepsilon \left( \mathscr{H}_- \tilde a_b^{-} + 2 \ii \beta \p_\eszeta \mathring a_b^{-} + \Sigma_b^{(-,0)}(\mathring a^{-})  +  \Sigma_b^{(-,2)}(\mathring a^{+})\right) &= 0\,,
\label{eq:15IV24.2b}
\end{align}
\end{subequations}
where $\Sigma_b^{(\pm, \Delta m)}$ are source terms introduced by the perturbation. The current formalism allows for arbitrary such $\Sigma_b^{(\pm, \Delta m)}$, explicit expressions for concrete models will be considered in Section \ref{sec:elasticity} below.

According to the multiple-scales method, both the leading-order terms and the $\varepsilon$-terms in \eqref{eq:15IV24.2} must vanish.
It follows from the analysis of the unperturbed problem that the zero-order solutions take the form
\begin{align}
	\label{eq:leading-order solution}
	\mathring a^\pm
		&= \mathscr J_\pm(\eszeta) \mathscr F_{\pm}(\mathring {\mathbf c}^\pm)\,,
\end{align}
which generalize the unperturbed solutions in \eqref{eq:15IV24.1} by allowing for a Jones vector $\mathscr J$ that depends on $\eszeta$.
Its components $\mathscr J_\pm(\eszeta)$ will be determined by the equations, as follows:
The $\varepsilon$-part of \eqref{eq:15IV24.2} is  a Helmholtz equation for $\tilde a_b^{\pm}$ with known source terms
\begin{align}
	\mathscr{H}_\pm \tilde a_b^{\pm} = - 2 \ii \beta \p_\eszeta \mathring a_b^{\pm} - \Sigma_b^{(\pm,0)} (\mathring a^{\pm})- \Sigma_b^{(\pm,2)} (\mathring a^{\mp})\,,
\label{eq:15IV24.3}
\end{align}
whose solutions are given by
\begin{align}
	 \tilde a^{\pm} = \mathscr{F}_\pm (\tilde{ \vecb{c}}^{\pm}) - \mathscr{G}_\pm \left(  2 \ii \beta \p_\eszeta \mathring a_b^{\pm} + \Sigma_b^{(\pm,0)} (\mathring a^{\pm})+ \Sigma_b^{(\pm,2)} (\mathring a^{\mp})\right)\,,
\label{eq:30VII24.1}
\end{align}
where $\mathscr{G}_\pm = \mathscr G_{\pm 1}$ is a multi-component operator \cite{2023PhRvR...5b3140M}
\begin{equation}
	\mathscr{G}_m (f) = \Big(
		G_m (f), \quad
		G_m (f), \quad
		G_{m+1} (f), \quad
		G_{m-1} (f)
	\Big)\,,
\end{equation}
in which $G_m (f)$ is the Green's operator for the Helmholtz equation
\begin{equation}
	G_m(f)
		:=
		\begin{cases}
			\frac{\pi}{2} Y_m (U r) \int_0^r\,  J_m (U s) f(s) s\, \dd s +  \frac{\pi}{2} J_m (U r) \int_r^\rho\,  Y_m (U s) f(s) s\, \dd s
				& r<\rho\,,
			\\
			- I_m (W r) \int_r^\infty \,  K_m (W s) f(s) s\, \dd s - K_m (W r) \int_\rho^r\,  I_m (W s) f(s) s \,\dd s
				& r>\rho\,.
		\end{cases}
\end{equation}
Here the $Y_m$'s are Bessel functions of the second kind, and $I_m$'s are modified Bessel functions of the first kind.

Eq.~\eqref{eq:30VII24.1} solves the field equations in the core and cladding separately.
To obtain a solution that is valid throughout the waveguide the field must satisfy the interface conditions given in Eq.~\eqref{eq:junction_conditions}.
In the perturbed case these can now contain functions of  order $\varepsilon$  which depend upon the radial and angular coordinates and the unperturbed solutions $\mathring a^\pm$. Following the reasoning leading to \eqref{eq:15IV24.2} we again restrict to isotropic and quadrupole perturbations, i.e., angular dependence $\Delta m  = \{0, \pm 2\}$, yielding the perturbed interface conditions
\begin{equation}\label{eq:30I25}
	\Gamma_\pm (\tilde a^\pm) + \esGamma^{(\pm,0)} (\mathring a^\pm) + \esGamma^{(\pm,2)} (\mathring a^\mp) = 0\,.
\end{equation}
Here $\Gamma_\pm$ is given in Eq.~\eqref{24VI24.1}, while $\esGamma^{(\pm,\Delta m)}$ are the correction terms specific to the perturbations under consideration and will be determined in Section~\ref{sec:elasticity}.

 These
 can be written  using \eqref{eq:30VII24.1} as
\begin{equation}
	M^\pm N^\pm \tilde{\vecb c}^{\pm}  = \Gamma_{\pm 1} \left[ \mathscr{G}_{\pm} \left(  2 \ii \beta \p_\eszeta \mathring a_b^{\pm} + \Sigma_b^{(\pm,0)} (\mathring a^{\pm}) +\Sigma_b^{(\pm,2)} (\mathring a^{\mp})\right) \right] -\esGamma^{(\pm,0)} (\mathring a^\pm) -\esGamma^{(\pm,2)} (\mathring a^\mp) \,,
\label{eq:15IV24.4}
\end{equation}
where $M^\pm = M_{\pm 1}$ and $N^\pm = N_{\pm 1}$ are given by Eqs.~\eqref{eq:matrix M} and \eqref{eq:matrix N}.
Since $M^\pm$ is singular due to the dispersion relation, the vectors given by the right-hand side of Eq.~\eqref{eq:15IV24.4} have to lie in the image of $M^\pm$. An explicit calculation shows that the cokernel of $M^\pm$ is one-dimensional. Let us denote by $\xi_\pm$ some basis vectors of the cokernel of $M^\pm$:
\begin{align}
	\xi_+ M^+ &= 0\,,
	&
	\xi_- M^- &= 0\,.
\end{align}
Then equation \eqref{eq:15IV24.4} admits solutions if and only if
\begin{equation}
	\xi_\pm \Gamma_\pm \left[ \mathscr{G}_\pm \left(  2 \ii \beta \frac{\p \mathring a_b^{\pm}}{\p \eszeta} + \Sigma_b^{(\pm,0)} (\mathring a^{\pm}) + \Sigma_b^{(\pm,2)} (\mathring a^{\mp})\right) \right] -\xi_\pm\esGamma^{(\pm,0)} (\mathring a^\pm) - \xi_\pm\esGamma^{(\pm,2)} (\mathring a^\mp)= 0\,.
\end{equation}
Since $\mathring a_b^+$ and $\mathring a_b^-$ depend on the coefficients $\mathscr J_+$ and $\mathscr J_-$, respectively, this equation provides the desired propagation law for the Jones vector.
Indeed, inserting \eqref{eq:leading-order solution} one obtains
\begin{align}
	\label{eq:Jones transport complex}
	\frac{\dd}{\dd \eszeta}
	\begin{pmatrix}
		\mathscr J_+\\
		\mathscr J_-
	\end{pmatrix}
	= \ii \mathcal M
	\begin{pmatrix}
		\mathscr J_+\\
		\mathscr J_-
	\end{pmatrix}\,,
\end{align}
where
\begin{align}
	\label{24VI24.2}
	\mathcal M
		&\equiv
		\begin{pmatrix}
			\t{\mathcal M}{_+^+} & \t{\mathcal M}{_+^-}\\
			\t{\mathcal M}{_-^+} & \t{\mathcal M}{_-^-}
		\end{pmatrix}
		\notag\\
		&= 
		\frac{1}{2\beta}
		\renewcommand{\arraystretch}{2.5}
		\begin{pmatrix}
				\displaystyle\frac{\xi_+ \left[\Gamma_+ \mathscr{G}_+ \Sigma^{(+,0)}(\mathscr F_{+})- \esGamma^{(+,0)}(\mathscr F_{+})\right]}{\xi_+ \Gamma_+ \mathscr{G}_+ (\mathscr F_{+})}
			&	\displaystyle\frac{\xi_+ \left[\Gamma_+ \mathscr{G}_+ \Sigma^{(+,2)}(\mathscr F_{-})-\esGamma^{(+,2)}(\mathscr F_{-})\right]}{\xi_+ \Gamma_+ \mathscr{G}_+ (\mathscr F_{+})}
			\\
				\displaystyle\frac{\xi_- \left[\Gamma_- \mathscr{G}_- \Sigma^{(-,2)}(\mathscr F_{+})-\esGamma^{(-,2)}(\mathscr F_{+})\right]}{\xi_- \Gamma_- \mathscr{G}_- (\mathscr F_{-})}
			&	\displaystyle\frac{\xi_- \left[\Gamma_- \mathscr{G}_- \Sigma^{(-,0)}(\mathscr F_{-})- \esGamma^{(-,0)}(\mathscr F_{-})\right]}{\xi_- \Gamma_- \mathscr{G}_- (\mathscr F_{-})}
		\end{pmatrix}
		\renewcommand{\arraystretch}{1.3}
		\,,
\end{align}
where $\mathscr F_{+}\equiv \mathscr F_{+}(\mathring {\mathbf c}^+)$ and $\mathscr F_{-}\equiv \mathscr F_{-}(\mathring {\mathbf c}^-)$.
The Cartesian components of the Jones vector thus satisfy
\begin{align}
	\label{eq:Jones transport cartesian}
	\frac{\dd}{\dd \eszeta}
	\begin{pmatrix}
		\mathscr J_x\\
		\mathscr J_y
	\end{pmatrix}
	=
	\ii \hat{\mathcal M}
	\begin{pmatrix}
		\mathscr J_x\\
		\mathscr J_y
	\end{pmatrix}\,,
\end{align}
where $\hat{\mathcal M}$ can be decomposed in terms of the four Pauli matrices as
\begin{align}
\begin{split}
	\hat{\mathcal M}
		={}& \half (\t{\mathcal M}{_+^+} + \t{\mathcal M}{_-^-}) \sigma_0
		+ \ihalf (\t{\mathcal M}{_+^-} + \t{\mathcal M}{_-^+}) \sigma_1
		\\
		+& \half (\t{\mathcal M}{_+^+} - \t{\mathcal M}{_-^-}) \sigma_2
		+ \half (\t{\mathcal M}{_+^-} + \t{\mathcal M}{_-^+}) \sigma_3\,.
\end{split}
\label{13I25.2}
\end{align}
In all cases below one has $\t{\mathcal M}{_+^+} = \t{\mathcal M}{_-^-}$ and $\t{\mathcal M}{_+^-} = \t{\mathcal M}{_-^+}$, so that $\hat{\mathcal M}$ takes to the form
\begin{align}
	\label{eq:matrix M simple}
	\hat{\mathcal M}
		= \t{\mathcal M}{_+^+}
		\begin{pmatrix}
			1	&	\hphantom{-}0\\
			0	&	\hphantom{-}1
		\end{pmatrix}
		+ \t{\mathcal M}{_+^-}
		\begin{pmatrix}
			1	&	\hphantom{-}0\\
			0	&	-1
		\end{pmatrix}\,.
\end{align}
In this case, the diagonal components of $\mathcal M$ (which arise from perturbations with $\Delta m = 0$) thus describe phase shifts, as they produce the same effect on $\mathscr J_x$ and $\mathscr J_y$, while off-diagonal components of $\mathcal M$ (corresponding to $\Delta m = \pm 2$) describe birefringence, producing opposite phase shifts on $\mathscr J_x$ and $\mathscr J_y$.

\section{Elasticity}
\label{sec:elasticity}

To continue, we wish to determine the source terms $\Sigma^{(\pm,0)}$ and $\Sigma^{(\pm,2)}$ coming from elastic perturbations of the fiber, after which \eqref{24VI24.2} can be evaluated numerically.

For this we consider  elastic deformations of the optical fibers resulting from ambient pressure, temperature, and gravity acting as a body force orthogonal to the principal axis of the fiber.
Considering a circular optical fiber as described in Sec.~\ref{sec:unperturbed}, we take the elastic parameters of the core and cladding to be identical and constant throughout the medium, resulting in a simple model of a homogeneous and isotropic cylinder of length $L$ and radius $a$.

In the following, we restrict the analysis to linear elasticity and  introduce the main equations, referring for a more detailed treatment to Ref.~\cite{BCS24}.
The displacement field $u_i$ describes small deformations of the material and determines the strain tensor $u_{ij}$ via
\begin{equation}\label{eq:17XII24.3}
	u_{ij} = \half (\p_i u_j + \p_j u_i)\,.
\end{equation}
The strain is related to the material stress $\sigma_{ij}$ via the generalized Hooke law for isotropic media \cite{Landau86,Sadd7}, as extended to account for changes of temperature $T-T_0$. Assuming the material to be thermally-linear and thermally-isotropic one has \cite{Sadd7}
\begin{equation}\label{eq:17XII24.2}
	\sigma_{ij} = \lambda u_{kk} \delta_{ij} + 2 \shearmod u_{ij} - (3\lambda + 2 \shearmod) \alphat (T-T_0) \delta_{ij}\,,
\end{equation}
where $\lambda$ and $ \shearmod$ are Lamé’s first and second parameters of the material, and where $\alphat$ is the coefficient of thermal expansion, all taken to be constant throughout the waveguide.
The second Lamé parameter $\shearmod$ is also referred to as the shear modulus, $\poisson = \lambda / [2 (\lambda + \shearmod)]$ is known as the Poisson ratio, and the Young modulus is given by $E = 2 \shearmod (1 + \poisson)$.
Material parameters for fused silica, which is widely used in optical fibers, are listed in Table~\ref{tab:SiO2}.
\begin{table}[ht]
	\centering
	\begin{tabular}{r|l}
			Poisson ratio $\nu$
		&	\num{0.17}
		\\
			Young modulus $E$
		&	\SI{73.1}{\giga\pascal}
		\\
			Mass density $\rho$
		&	\SI{2.2}{\gram\per\cubic\centi\meter}
		\\
			Thermal expansion coefficient $\alpha$
		&	\SI{5.5e-7}{\per\kelvin}
	\end{tabular}
\caption{Properties of silica oxide glass from Ref.~\cite{Crystran12}.}
\label{tab:SiO2}
\end{table}

Static configurations are described by solutions to the equilibrium equations \cite{Sadd7}
\begin{equation}\label{eq:eq_eqs}
	\p_j \sigma_{ij} + F_i = 0\,,
\end{equation}
with $F_i = - \p_i V$ denoting the gravitational body force, where $V = \grav \uprho y$ is the gravitational potential, with $\uprho$ being the mass density of the material and $\grav$ the gravitational acceleration of Earth.
At the object’s surface, one has the traction boundary conditions \cite{Landau86}
\begin{equation}\label{18XI24.1}
	\sigma_{ij} n_j = P_i\,,
\end{equation}
where $n_j$ is the outward-pointing unit normal vector of the material surface and $P_i$ is the external pressure.

Taking into account the translational symmetry along the fiber, as well as the fact that the  radius $a$ of the fiber is  significantly smaller than its length $L$, $a \ll L$, the calculations can be reduced to those on two-dimensional cross-sections orthogonal to the main axis.
The textbook approach does so by setting the out-of-plane strain components to zero \cite[Chapter 7]{Sadd7} leading to algebraic expressions for the out-of-plane stress and two-dimensional equilibrium equations \eqref{eq:eq_eqs}.
This can be generalized by allowing for a linear displacement along the fiber (cf. \cite{Zhenye90})
\begin{equation}
	u_z = \kappa z\,,
\end{equation}
where $\kappa$ is a constant, implying
\begin{align}
	&u_{xz}=u_{yz} = 0\,,
	&
	\text{and}
	&
	&
	u_{zz} = \kappa\,.
\end{align}
The equilibrium equations can now be solved by introducing a stress function $\phi$, related to the stress tensor in cylindrical frame components via
\begin{align}\label{eq:17XII24.1}
	\sigma_{rr} &= \frac{1}{r} \p_r \phi + \frac{1}{r^2} \p_\theta^2 \phi + V\,,
	&
	\sigma_{r\theta} &=  \p_r^2 \phi + V\,,
	&
	\sigma_{\theta\theta} &= -\p_r \left(\frac{1}{r} \p_\theta \phi\right)\,.
\end{align}
The general solution to these equations has been written down by Michell \cite{Michell}.
Assuming regularity, vertical reflection symmetry, and absence of friction at the fiber’s boundary one is led to \cite{BCS24}
\begin{equation}\label{eq:14I25.12}
	\phi (r,\theta) = d_0 r^2  + \sum_{\ell\geq 2} \left( 1 +  \frac{1-\ell}{1+\ell} \frac{r^2}{ a^2}\right) b_\ell r^\ell\cos \left[\ell\left(\theta+\frac{\pi}{2}\right)\right]\,,
\end{equation}
where the angle $\theta$ used here has been shifted by $\pi/2$ relative to that in \cite{BCS24} to measure the usual angle from the horizontal.
The parameters $d_0$ and $b_\ell$ are determined by the boundary condition \eqref{18XI24.1}.
Combining \eqref{eq:17XII24.2} and \eqref{eq:17XII24.1} yields a parameterized expression for the strain tensor,
\begin{subequations}\label{eq:17XII24.4}
\begin{align}
	\begin{split}
		u_{rr} ={}
			& - \frac{1}{2\shearmod} \Big\{ - (1-2\poisson)\grav r \uprho \sin(\theta) - 2 (1-2\poisson) d_0 + 2 \shearmod \poisson \alpham - 2 \shearmod(1+ \poisson) \alphat (T-T_0)
			\\&\hspace{3em}
			- \sum_{\ell=2}^\infty \left[ \ell  -   (\ell-2 + 4 \poisson) \frac{r^2}{a^2} \right]  (1-\ell)r^{\ell-2} b_\ell \cos \left[\ell\left(\theta+\frac{\pi}{2}\right)\right]
			 \Big\}\,,
	\end{split}\label{eq:16XII24.1a}
	\\
	\begin{split}
		u_{r\theta} ={}
			&- \frac{1}{2 \shearmod}\sum_{\ell = 2}^{\infty} \left[ 1 -  \frac{r^2}{a^2}\right] (1-\ell) \ell r^{\ell-2} b_\ell\sin \left[\ell\left(\theta+\frac{\pi}{2}\right)\right]\,,
	\end{split}
	\\
	\begin{split}
		u_{\theta\theta} ={}
			& - \frac{1}{2\shearmod} \Big\{ - (1-2\poisson)\grav r \uprho \sin(\theta) - 2 (1-2\poisson) d_0  + 2 \shearmod \poisson \alpham - 2 \shearmod(1+ \poisson) \alphat (T-T_0)
			\\&\hspace{3em}
			+ \sum_{\ell=2}^\infty \left[  \ell  -  (\ell+2 - 4 \poisson)  \frac{r^2}{a^2} \right](1-\ell)r^{\ell-2} b_\ell\cos \left[\ell\left(\theta+\frac{\pi}{2}\right)\right]
			 \Big\}\,.
	\end{split}\label{eq:16XII24.1c}
\end{align}
\end{subequations}
This can be integrated using \eqref{eq:17XII24.3}, leading to the following form of the frame-components of the displacement vector:
\begin{subequations}\label{eq:15II24.2}
\begin{align}
	\label{eq:15II24.2a}
	\begin{split}
		u_r (r, \theta)
			={}& \shearmod^{-1}(1 - 2 \nu) d_0  r
			- \poisson \alpham  r
			+ (1 + \poisson )\alphat r (T-T_0)
			\\&
			+ \frac{1}{2 \shearmod}\Big\{
				\left[
					 \half \grav \uprho (1 - 2 \nu) r^2
					+ 2 \shearmod \Xi
				\right] \sin(\theta)
			\\&\hspace{3em}
				- \sum_{\ell \geq 2} \left[
					 \ell
					-(2 - 4\nu - \ell ) \frac{1-\ell}{1+\ell} \frac{r^2}{a^2}
				\right] b_\ell  r^{\ell -1} \cos \left[\ell\left(\theta+\frac{\pi}{2}\right)\right]
			\Big\}\,,
	\end{split}
	\\
	\label{eq:15II24.2b}
	\begin{split}
		u_\theta (r, \theta)
			={}&
			\frac{1}{2 \shearmod} \Big\{
				\left[
					- \half \grav \uprho (1 - 2 \nu) r^2
					+ 2 \shearmod \Xi
				\right] \cos (\theta)
			\\&\hspace{3em}
			+ \sum_{\ell  \geq 2} \left[
					\ell
					+ (4 - 4\nu + \ell ) \frac{1-\ell}{1+\ell} \frac{r^2}{a^2}
				\right] b_\ell  r^{\ell -1} \sin \left[\ell\left(\theta+\frac{\pi}{2}\right)\right]
			\Big\}\,.
	\end{split}
\end{align}
\end{subequations}
The parameter $\Xi$ describes rigid vertical displacements of the fiber and can thus be determined, for example, by prescribing the displacement at a point of the fiber’s boundary, e.g., by setting $u_r (a,-\tfrac{\pi}{2}) = u_\theta (a,-\tfrac{\pi}{2}) = 0$. 

Anticipating, the only expansion coefficients of the Michell solution that couple to the optical phase and polarization are $d_0$ and $b_2$.
It is worth noting that these two coefficients are completely determined by the strain tensor at the axis of the fiber. Indeed, in Cartesian coordinates we have
\begin{subequations}\label{14I15.11}
	\begin{align}
		\Kk &= 2 \shearmod^{-1}(1-2\poisson) d_0 - 2\poisson \alpham + 2(1+\poisson) \alphat (T-T_0)\,,
		\\
		\Bb &= -2 \shearmod^{-1} b_2\,.
	\end{align}
\end{subequations}

In Ref.~\cite{BCS24} a number of physically motivated boundary conditions were considered. For the numerics of Section~\ref{sec:numerics} we restrict ourselves to the model discussed in Ref.~\cite[Section 4.2.1]{BCS24}, namely the waveguide resting on a rigid plane with an extended contact region. In this case one has
\begin{align}
	d_0 &= - \frac 14 \left( a \grav \rho + 2 \mathfrak p \right)\,,
	&
	b_\ell &= \frac{\grav \rho  }{2 a^{\ell - 3} (\ell - 1)}\,,
	\quad \ell \geq 2 \,,
\end{align}
where $\mathfrak p$ is an ambient isotropic pressure.

The following sections \ref{sec:displacement} and \ref{sect:photoelasticity} describe the influence of such fiber deformations on the electromagnetic modes propagating therein.

\subsection{Displacement}\label{sec:displacement}
 
Transverse deformations of an optical fiber which are invariant under translation along its axis are described by displacements of the form
\begin{align}\label{eq:15II24.1}
	r &\rightarrow r + u_r (r, \theta)\,,
	&
	\theta &\rightarrow \theta + u_\theta (r, \theta) / r\,,
\end{align}
which is assumed to be sufficiently small so that nonlinear terms in $u_r$ and $u_\theta$ are negligible. 

The deformation implies that the core-cladding interface is displaced from its reference location, while the calculations in Sect.~\ref{sec:multiplescale} require the core-cladding interface to lie at a constant radial coordinate distance from the fiber’s axis. This can be resolved by performing a coordinate transformation, namely the inverse of \eqref{eq:15II24.1}, so that the core-cladding interface is again located at $r = \rho$ in the new coordinates.
This coordinate transformation results in additional source terms in the field equations, similarly for the interface equations.

The details are as follows.

\paragraph{The equations.}
For definiteness let 
\begin{align}
	(x^\mu) \equiv (t, \vec x )\equiv (t, x^i)\equiv (t,x,y,z)
\end{align}
denote the coordinates before the waveguide has been deformed, and let $(\bx^\mu) \equiv (t, \bx^i)\equiv (t,\bx,\by,\bz)$ describe the deformed configuration:
\begin{equation}\label{21I25.1}
  \bx^i = x^i + u^i(\vec x)
  \,.
\end{equation} 
The interface between the core and the cladding is located at $x^2 + y^2 = \rho^2$.

We assume that $u^i$ and  its derivatives  are of order $\varepsilon$, where $\varepsilon$ is small compared to the remaining scales involved. In particular the map $x^i\mapsto \bx^i$ is invertible with
\begin{align}\label{21I25.5}
	x^i &= \bx^i +  O(\varepsilon)\,,
	&
	x^i &= \bx^i - u^i|_{x^j=\bx^j} +  O(\varepsilon^2) \,.
\end{align} 
Denoting by $\bA_\mu$ the vector potential in the barred coordinate system,
\begin{equation}\label{21I25.2}
	\bA_\mu (t,\bx,\by,\bz)|_{\bx^i=x^i+u^i(\vec x)}
	= A_\nu(t,\vec x) \frac{\partial x^\nu}{\partial \bx^\mu }\,,
\end{equation}
one has
\begin{align}\label{21I25.2a}
	\bA_0
		&= A_0 
		- \frac{\partial A_0}{\partial x^i} u^i 
		+ O(\varepsilon^2)\,,
	&
	\bA_i
		&= A_i
		- \frac{\partial A_i}{\partial x^j } u^j
		- A_j  \frac{\partial u^j}{\partial x^i }
		+ O (\varepsilon^2)\,.
\end{align}
Away from the interface the vector potential  $\bA_\mu$ satisfies the wave equation
\begin{align}
	\label{10VI24.1a}
	(- n^2 \partial_t^2 + \partial^2_\bx + \partial^2_\by + \partial^2_\bz) \t 
 \bA{_\mu} = 0\,.
\end{align}
In the $x^\mu$-coordinates this becomes
\begin{align}
	\label{10VI24.1b}
	(- n^2 \partial_t^2 +g^{ij}\nabla_i\nabla_j) \t  A{_\mu} = 0\,,
\end{align}
where $g^{ij}$ is the space-part of the transformed contravariant metric
\begin{align}
	\label{21I25.3}
	g^{\mu\nu}  (t, x,y,z)
= \Big(\bg^{\alpha\beta}
  \frac{\partial x^\mu}{\partial \bx^\alpha }
  \frac{\partial x^\nu}{\partial \bx^\beta}
  \Big)
   \Big|_{\bx^i=x^i+u^i(\vec x)}  
 \equiv \Big(\eta^{\alpha\beta}
  \frac{\partial x^\mu}{\partial \bx^\alpha }
  \frac{\partial x^\nu}{\partial \bx^\beta}
  \Big)
   \Big|_{\bx^i=x^i+u^i(\vec x)}
   \,,
\end{align}
with $\nabla_\mu$ denoting the covariant derivative of the transformed metric. In particular, keeping in mind that the $\bx^i$'s are Euclidean coordinates in which $\bg^{ij} = \delta^{ij} = \delta_{ij}$ where $\delta^i_j$ is the Kronecker delta, 
\begin{align}
	\label{21I25.4} 
	g^{ij}(x,y,z)
		= \Big(\delta^{k\ell}
			\frac{\partial x^i}{\partial \bx^k}
			\frac{\partial x^j}{\partial \bx^\ell}
		\Big) \Big|_{\bx^i=x^i+u^i(\vec x)}
		= \delta^{ij}
		- \partial_i u^j
		- \partial_j u^i
		+ O(\varepsilon^2)
		= \delta^{ij} + O(\varepsilon )\,.
\end{align}
The Christoffel symbols are 
\begin{align}
  & \Gamma^0_{\alpha\beta} = \Gamma^\alpha_{0\beta} = 0\,,
  &
  & \Gamma^i_{jk} = \partial_j \partial_k u^i + O(\varepsilon^2) = O(\varepsilon)\,,
\end{align}
so that
\begin{subequations}
\begin{align}
	& \nabla_0 A_0
		= \partial_0 A_0\,,
	\qquad\quad
	\nabla_i A_0
		= \partial_i A_0 \,,
	\qquad\quad
	\nabla_i \nabla_j A_0
		= \partial_i \partial_j A_0
		- (\t\p{_i} \t\p{_j} \t u{^k}) \t\p{_k} \t A{_0}
		\,,
	\\
	& \nabla_j A_k
		= \partial_j A_k
		- A_\ell \partial_j \partial_k u^\ell
		+ O(\varepsilon^2)\,,
	\\   
	& \nabla_i\nabla_j A_k
		= \partial_i \big(\partial_j A_k -   A_\ell \partial_j \partial_k u^\ell \big)
		- \partial_i \partial_j u^\ell  \partial_\ell A_k 
		- \partial_i \partial_k u^\ell  \partial_j A_\ell 
		+ O(\varepsilon^2)
   \,,
\end{align}
\end{subequations}
Hence, up to terms of order $\varepsilon^2$, the system \eqref{10VI24.1b} can be rewritten as
\begin{subequations}
\begin{align}
	\label{21I25.7}
 	(- n^2 \partial_t^2 +\delta ^{ij}\partial_i  \partial_j) \t A{_0}
 	= \ &
   		\delta^{ij}\left[2\p_i u^\ell\partial_j  \partial_\ell \t  A{_0}
		+  \t\p{_i} \t\p{_j} \t u{^\ell} \t\p{_\ell} \t A{_0}\right]
		\,,
	\\
	\begin{split}
		(- n^2 \partial_t^2 +\delta^{ij}\partial_i  \partial_j) \t A{_k}
		  = \ &
		 \delta^{ij}\left[ 2\p_i u^\ell\partial_j  \partial_\ell \t  A{_k}
				+ \partial_i \partial_j u^\ell  \partial_\ell A_k 
				+ 2\partial_i \partial_k u^\ell  \partial_j A_\ell 
				 + A_\ell \partial_i\partial_j \partial_k u^\ell
			\right]\,.
	\end{split}
\end{align}
\end{subequations}

\subsubsection{Boundary conditions}
 \label{ss23I25.1}

Let $n^i$ denote the field of normals to the interface between the core and the cladding. In the unbarred coordinate system the interface is located at $r\equiv \sqrt {x^2 + y^2}=\rho$, 
where $\vec x$ denotes the position vector in the plane perpendicular to the axis of the fiber. The covariant components of the normal  $n_A$, $A=1,2$, are proportional to $x^A/r$, say $n_A=\alpha x^A/r$.
As already mentioned above, the normalization of $n_A$ is irrelevant for the equations, and thus we take
  \begin{eqnarray}
 	\label{eq:conormal normalized}
   n_A
    &  =  & 
 	  \frac{x^A}{r}  + O(\wepsilon^2) 
 \,,
 \qquad 
  n_z=0
 	 \,.
  \end{eqnarray}
As in the unperturbed case, the matching conditions at the core-cladding interface take the general form \eqref{eq:junction_conditions}, reproduced here for the convenience of the reader
\begin{align}
	\llbracket \chi_\text{gauge} \rrbracket &= 0\,,
	&
	\llbracket \t A{_\mu} \rrbracket &= 0\,,
	&
	\llbracket \t G{^\mu^\nu} \rrbracket n{_\nu} &= 0\,,
\end{align}
with
$$
 (n_\mu )
 := (n_0,n_i)  \equiv (n_t,n_i) = (0,n_1,n_2,0)
 \,.
$$
Here, as above, $\llbracket f\rrbracket  = (f_\text{core} - f_\text{cladding}) |_{r=\rho}$ denotes the jump of a function $f$ at the surface where $r^2 = x^2 + y^2$ equals $\rho^2$. 

To continue, we need explicit formulae for $\t G{^\mu^\nu}n{_\nu}$. Keeping in mind our assumption that $\permeability = 1$, we have
\begin{subequations}
\begin{align} 
\notag
  \t G{^t^A}n_{A}
  = & \ \t G{^t^r} =  \t \gamma{^t^t}\t \gamma{^r^\nu}(\p_t A_\nu - \p_\nu A_t)
  \\  \notag
  = &\ - n^2(1- 2\p_r u_r)(\p_t A_r - \p_r A_t) 
  \\ &\ - n^2 r^{-2} (u_\theta - \p_\theta u_r - r \p_r u_\theta) (\p_t A_\theta - \p_\theta A_t)  + O(\wepsilon^2)
   \,,
\\     
 \t G{^r^A}n{_A}
  =& \ \t G{^r^r} = 0
   \,, 
   \\     
   \notag
 \t G{^\theta^A}n{_A}
  =& \ \t G{^\theta^r} =   \t \gamma{^\theta^r}\t \gamma{^r^\theta}(\p_r A_\theta - \p_\theta A_r)  +  \t \gamma{^\theta^\theta}\t \gamma{^r^r}(\p_\theta A_r - \p_r A_\theta) 
  \\
  = &\ r^{-2} \left[ 1 - 2 r^{-1} (u_r +\p_\theta u_\theta +  \half r \p_r u_r)\right] (\p_\theta A_r - \p_r A_\theta) + O(\wepsilon^2)
   \,, 
\\     
\notag
\t G{^z^A}n{_A}
  =& \ \t G{^z^r} =  \t \gamma{^z^z}\t \gamma{^r^\nu}(\p_z A_\nu - \p_\nu A_z)
  \\\notag
  = &\ (1- 2\p_r u_r)(\p_z A_r - \p_r A_z) 
  \\ &\ + r^{-2} (u_\theta - \p_\theta u_r - r \p_r u_\theta) (\p_z A_\theta - \p_\theta A_z)  
  + O(\wepsilon^2)
   \,.
\end{align}
\end{subequations}
Thus, assuming that  
\begin{align}
	\llbracket \t A{_\mu} \rrbracket &= 0\,, 
\end{align}
we obtain
\begin{subequations}
\begin{align}\label{30I25.1}
  \llbracket
  \t G{^t^\nu}n{_\nu}
  \rrbracket 
  = 
   & \
    -  \left(
 	 \frac{x^B}r  - \partial_r u^B
 	 - \partial_B u^r 
 	 \right) 
 	 \llbracket n^2
  (\partial_t A_B-\partial_B A_t) \rrbracket   + O(\wepsilon^2)
   \,,
\\   
 \llbracket  
 \t G{^r^\nu}n{_\nu}
 \rrbracket 
  = 
   & \ \llbracket  G^{rr}\rrbracket +  O(\wepsilon^2) = O (\wepsilon^2)\,,
   \\     
 \llbracket  
 \t G{^\theta^\nu}n{_\nu}
 \rrbracket 
  = 
   & \ r^{-2} \left[ 1 - 2 r^{-1} (u_r +\p_\theta u_\theta + \half r \p_r u_r)\right]  \llbracket  
\p_r A_\theta
 \rrbracket  + O(\wepsilon^2)    
   \,, 
\\   
 \llbracket
 \t G{^z^\nu}n{_\nu}
 \rrbracket 
  = 
   & \ 
    (1 - 2\partial_r u^r) 
  \llbracket \partial_r A_z \rrbracket  + O(\wepsilon^2)
   \,,
\end{align}
\end{subequations}
and
\begin{equation}
	\llbracket \chi_\text{gauge}\rrbracket =\llbracket -  n^2\p_t A_0    + (1- 2\p_r u_r)  \p_r  A_r + r^{-2}  (u_\theta - \p_\theta u_r - r \p_r u_\theta) \p_r  A_\theta \rrbracket+ O(\wepsilon^2) \,.
\end{equation}

\subsubsection{Explicit expressions}

As already pointed-out in Sect.~\ref{sec:multiplescale}, the results obtained from first-order perturbation theory only depend on perturbation terms with relative azimuthal mode number $\Delta m \in \{0, +2, -2\}$, i.e., only terms without angular dependence or $\propto \ee^{\pm 2 \ii \theta}$.
We treat these cases independently in the following paragraphs.

\paragraph{Radial perturbations}
In the simplest case $\Delta m = 0$ the perturbative terms are independent of the angular coordinate $\theta$, and we can write the wave equation again in terms of Helmholtz operators
\begin{equation}
	(\mathscr{H}_m + \varepsilon \delta \mathscr{H}_m) a_b = 0
 \,.
  \label{13I25.1}
\end{equation}
Explicitly, for the displacements given in \eqref{eq:15II24.1} together with \eqref{eq:15II24.2} one finds
\begin{equation}
	\varepsilon \delta \mathscr{H}_m = -2 \left( \shearmod^{-1}(1- 2 \poisson) d_0 - \poisson \alpham + (1+\poisson) \alphat (T - T_0)\right) ( n^2 \omega^2 - \beta^2)\,.
\end{equation}
This is equivalent to the unperturbed equation \eqref{10II24.1} with rescaled frequency and wave vector. Consequently, the solutions are given as in \eqref{eq:15II24.3}–\eqref{eq:15IV24.1} with the substitutions
\begin{subequations}\label{eq:30VI24.1ab}
\begin{align}\label{eq:30VI24.1}
	U &\rightarrow [1 - (1- 2 \poisson) d_0 / \shearmod+ \poisson \alpham- (1+\poisson) \alphat (T - T_0)] U\,,
		\\
	W &\rightarrow [1 -(1- 2 \poisson) d_0 / \shearmod + \poisson \alpham - (1+\poisson) \alphat (T - T_0)] W\,.
	 \label{eq:30VI24.2}
\end{align}
\end{subequations}
This, in turn, introduces a shift in the roots of \eqref{15II24.4} and thus a perturbation of $\beta$, namely
\begin{align}
	\frac{\delta \beta}{\beta}
		&= [(1 - 2 \poisson) d_0 / \shearmod - \poisson \alpham + (1 + \poisson) \alphat (T - T_0)] V \frac{n_1^2 - n_2^2}{2 \bar n^2} \frac{\dd b}{\dd V}\,,
\end{align}
where $\bar n = \beta / \omega = \sqrt{ b n_1^2 + (1-b) n_2^2 }$ is the effective refractive index, and $\dd b / \dd V$ is the derivative of the normalized guide index with respect to the normalized frequency, which can be read from Fig.~\ref{fig:unperturbed_modes}.

Alternatively, in the current simple case one can directly work in the original coordinates, with $\delta \mathscr{H}_m =0$ in \eqref{13I25.1}. The interface region is then displaced as $\rho \rightarrow \rho + u_r (\rho)$. Such a perturbation in Eq.~\eqref{eq:UW} is equivalent to \eqref{eq:30VI24.1ab}.

\paragraph{Perturbations with angular dependence}
Apart from the angle-independent terms described above, the displacements given in Eqs.~\eqref{eq:15II24.2a}--\eqref{eq:15II24.2b} lead to correction terms to the field equations with angular dependence $\ee^{\pm 2 \ii \theta}$ that can be written in the form of Eqs.~\eqref{eq:15IV24.2a}--\eqref{eq:15IV24.2b} with
\begin{subequations}
\begin{align}
	\begin{split}
		\varepsilon \Sigma_t^{(\pm,2)}
			&= \frac{b_2}{\mu r^2 a^2} \Big[
					(a^2+2(1-\poisson)r^2) \mathring a^{\mp}_t
					- r( a^2 - 2 r^2 (1+\poisson)) \p_r \mathring a^{\mp}_t
					\\&\hspace{5em}
					- r^2 (a^2-2r^2\poisson)\p_r^2 \mathring a^{\mp}_t
				\Big]\,,
	\end{split}
	\\
	\begin{split}
		\varepsilon\Sigma_z^{(\pm,2)}
			&= \frac{b_2}{\mu r^2 a^2} \Big[
				(a^2+2(1-\poisson)r^2) \mathring a^{\mp}_z
				- r( a^2 - 2 r^2 (1+\poisson)) \p_r \mathring  a^{\mp}_z
				\\&\hspace{5em}
				- r^2 (a^2-2r^2\poisson)\p_r^2 \mathring a^{\mp}_z
			\Big]\,,
	\end{split}
	\\
	\begin{split}
		\varepsilon\Sigma_\sharp^{(+,2)}
			&= \frac{b_2}{\mu r a^2} \Big[
				(a^2+2r^2(1+\poisson))\p_r \mathring a_\sharp^{-}
				- r (a^2 - 2 r^2 \poisson)\p_r^2  \mathring a_\sharp^{-}
			\Big]\,,
	\end{split}
	\\
	\begin{split}
		\varepsilon\Sigma_\flat^{(-,2)}
			&= \frac{b_2}{\mu r a^2} \Big[
				(a^2+2r^2(1+\poisson))\p_r \mathring a_\flat^{+}
				- r (a^2 - 2 r^2 \poisson)\p_r^2  \mathring a_\flat^{+}
			\Big]\,,
	\end{split}
	\\
	\begin{split}
		\varepsilon\Sigma_\flat^{(+,2)}
			&= \frac{b_2}{\mu r a^2} \Big[
					2 r (-2 + 4 \poisson) \mathring a_\flat^{-}
					- (3a^2 + 2 r^2 (1 - 5 \poisson)) \p_r \mathring a_\flat^{-} - r (a^2 - 2 r^2 \poisson) \p_r^2 \mathring a_\flat^{-}
					\\&\hspace{5em}
					+ 4 r \mathring a_\sharp^{-}
					+ 4 r^2 \p_r \mathring a_\sharp^{-}
			\Big]\,,
	\end{split}
	\\
	\begin{split}
		\varepsilon\Sigma_\sharp^{(-,2)}
			&= \frac{b_2}{\mu r a^2} \Big[
					2 r (-2 + 4 \poisson) \mathring  a_\sharp^{+}
					- (3a^2 + 2 r^2 (1 - 5 \poisson)) \p_r \mathring a_\sharp^{+} - r (a^2 - 2 r^2 \poisson) \p_r^2 \mathring a_\sharp^{+}
					\\&\hspace{5em}
					+ 4 r \mathring a_\flat^{+}
					+ 4 r^2 \p_r \mathring a_\flat^{+}
			\Big]\,.
	\end{split}
\end{align}
\end{subequations}
For the correction terms arising for the interface conditions \eqref{eq:30I25} we find explicitly
\begin{eqnarray}
	\esGamma^{(\pm,2)} [\mathring a^\mp] =
	\ii b_2 \shearmod^{-1} 
	\begin{pmatrix}
	 	0 \\
		0 \\
	 	0 \\
	  	0 \\
		\pm   (1-\frac{\rho^2}{a^2} ) \left\llbracket    n^2  \left(\mp \ii r^{-1}\mathring a_t^\mp - \frac{1}{\sqrt 2}\omega \mathring a_\sharp^\mp + \frac{1}{\sqrt 2}\omega \mathring a_\flat^\mp\right)\right\rrbracket \\
		0\\
		0 \\
		-\omega  \left(1- 2 \poisson \frac{\rho^2}{a^2}\right) \left\llbracket  n^2 \mathring a_t^\mp \right\rrbracket\\
	\end{pmatrix}\,.
\end{eqnarray}
These expressions can be used in Eq.~\eqref{24VI24.2} to compute the matrix $\mathcal M$ to obtain an explicit form of the propagation law given by Eqs.~\eqref{eq:Jones transport complex}-\eqref{24VI24.2}; equivalently
 \eqref{eq:Jones transport cartesian}-\eqref{13I25.2}.
Since the dispersion relation can only be solved numerically and the integrals arising here admit no concise explicit form, we evaluate $\mathcal M$ numerically.
Concrete numerical results for a range of fiber parameters, including those relevant for the GRAVITES experiment \cite{Hilweg:2017ioz}, are presented in Section \ref{sec:numerics} and Figures~\ref{fig:birefringence}--\ref{fig:phase shifts}.

\subsection{Photoelasticity}
\label{sect:photoelasticity}

We introduced the constitutive tensor for linear dielectric media in \eqref{18V24.1},   specializing so far to isotropic media.
Now, we consider small deviations from an ideal, isotropic, non-magnetic dielectric, writing the constitutive tensor as
\begin{equation}
	\chi^{\mu\nu\rho\sigma} = \chi^{\mu\nu\rho\sigma}_\text{isotropic}+ \varepsilon \delta \chi^{\mu\nu\rho\sigma}\,,
\end{equation}
where $\varepsilon$ is the same small expansion parameter introduced in \eqref{eq:perturbed wave equation}, with a tensor field $ \delta \chi^{\mu\nu\rho\sigma}$ arising from the change in permittivity introduced by elastic deformations.
Taking a perturbative expansion for the field strength tensor $F_{\mu\nu}$ as well, we can write \eqref{18V24.1} as
\begin{equation}
	G^{\mu\nu} =   \gamma^{\mu\rho}\gamma^{\nu\sigma} \left(\mathring F_{\rho\sigma} + \varepsilon\tilde F_{\rho\sigma} \right) + \varepsilon \delta \chi^{\mu\nu\rho\sigma} \mathring F_{\rho\sigma}  \,,
\end{equation}
so that, using \eqref{13I25.4}, the gauge-fixed equation \eqref{eq:maxwell + gauge} becomes
\begin{equation}\label{eq:30VII24.2}
 \Box_\gamma  A_\beta
= - 2 \gamma_{\beta \nu}\nabla_\mu (\varepsilon \delta \chi^{\mu\nu\rho\sigma} \p_{\rho} \mathring A_\sigma) + O(\varepsilon^2)\,.
 \end{equation}
We assume  that the medium remains non-magnetic and induces no magneto-electric coupling, so that $\delta \chi^{\mu\nu\rho\sigma}$ takes the form
\begin{align}
	\varepsilon\delta \chi^{0i0j} &= \delta \permittivity^{ij}\,,
	&
	\delta \chi^{ijkl} &= 0\,,
	&
	\delta \chi^{0ijk} &= 0\,,
\end{align}
where $\delta \permittivity^{ij}$ is the elastically induced change in the permittivity tensor.
In the linear regime, $\delta \permittivity^{ij}$ is related to the material strain $u_{kl}$ via the photoelasticity tensor $p_{ijkl}$ \cite[Appendix D]{chen06}:
\begin{equation}
	\delta (\permittivity^{-1})_{ij} = p_{ijkl} u_{kl}\,.
\end{equation}
The components $p_{ijkl}$ generally need to be determined empirically, though in the case of an isotropic solid they simplify to \cite{narasimhamurty81}
\begin{equation}\label{eq:isotropic_photoelasticity_tensor}
	\vecb p =
\begin{pmatrix}
	p_{11} & p_{12} & p_{12} & 0 & 0 & 0 \\
	p_{12} & p_{11} & p_{12} & 0 & 0 & 0 \\
	p_{12} & p_{12} & p_{11} & 0 & 0 & 0 \\
	0	&	0	&	0	&  \tfrac 12 (p_{11} - p_{12}) & 0 & 0 \\
	0	&	0	&	0	&	0	&  \tfrac 12 (p_{11} - p_{12}) & 0 \\
	0	&	0	&	0	&	0	&	0	&  \tfrac 12 (p_{11} - p_{12}) \\
\end{pmatrix}
\,,
\end{equation}
in Voigt notation, where $11 \rightarrow 1, 22 \rightarrow 2, 33 \rightarrow 3, 23 \rightarrow 4, 13 \rightarrow 5, 12 \rightarrow 6$.
Alternatively, employing tensor notation, we introduce material constants $\Pp$ and $\Qq$ through the formula
\begin{align}
	\begin{split}
		p_{ijkl}
			&= \Pp g_{ij}g_{kl}
			+ \Qq (g_{ik}g_{jl} + g_{il}g_{jk})
		\\&
			\equiv p_{12} g_{ij}g_{kl}
			+ \half (p_{11}-p_{12}) (g_{ik}g_{jl} + g_{il}g_{jk})
		\,,
	\end{split}
\end{align}
where $\t g{_i_j}$ are the components of the spatial metric (equal to the Kronecker  $\t\delta{_i_j}$ in Cartesian coordinates).
For the case of fused silica glass, as used in optical fibers, the photoelasticity tensor is characterized by \cite{Biegelsen74,Primak59}
\begin{align}
	p_{1111} &\equiv p_{11} = 0.121\,,
	&
	\Pp \equiv p_{1122} &\equiv p_{12} = 0.271\,,
	&
	\Qq \equiv p_{1212} &\equiv p_{33} = -0.075\,.
\end{align}
Note that $p_{1212}$ is measured independently, but in good agreement with the isotropy assumption made in \eqref{eq:isotropic_photoelasticity_tensor}.
The induced source terms for the perturbative calculation as determined via \eqref{eq:30VII24.2} and the strain tensor \eqref{eq:17XII24.4} are given by
\begin{subequations}
\begin{align}
\notag	\varepsilon \Sigma_t^{(\pm,0)} &=- n^2\frac{\Kk}{  r^2} \Big\{
	2 (\Pp + \Qq + \Pp r^2 \beta^2) a_t^\pm + r \Big(\pm  \ii \sqrt{2} (\Pp + \Qq) \omega a_\flat^\pm 
	\\&\hspace{1em} \mp  \ii \sqrt 2 (\Pp + \Qq) \omega a_\sharp^\pm + 2\Pp r \beta \omega a_z^\pm -  \ii (\Pp + \Qq) r [ \sqrt 2 \omega \p_r( a_\flat^\pm + a_\sharp^\pm) - 2 \ii \p_r^2 a_t^\pm]
	\Big)
	\Big\}
\,,\label{eq:16XII24.2a}
\\
	\varepsilon\Sigma_z^{(\pm,0)}  &=  \Kk 2  n^4 \omega \Pp \left(  \beta a_t^\pm + \omega a_z^\pm\right)
\,,
\\
	\varepsilon\Sigma_\sharp^{(\pm,0)}  &=  \frac{ \Kk}{ r} (\Pp + \Qq) n^4  \omega \left( 2 r \omega a_\sharp^\pm \pm \ii \sqrt 2 \left[ a_t^\pm \mp r \p_r a_t^\pm \right]\right)
\,,
\\	\varepsilon\Sigma_\flat^{(\pm,0)} &=   \frac{ \Kk }{r} (\Pp + \Qq)  n^4 \omega \left( 2 r \omega a_\flat^\pm \mp \ii \sqrt 2 \left[ a_t^\pm \pm r \p_r a_t^\pm \right]\right)
\,,\label{eq:16XII24.2d}
\end{align}
\end{subequations}
as well as
\begin{subequations}
\begin{align}
\notag	\varepsilon \Sigma_t^{(+,2)} =& -n^2 \frac{\Bb}{4  r^2 a^2} \times
		\\\notag&  \Big\{
		4\left[\Qq r^2 (1-2\poisson) + \Pp r^2 (-1 + r^2 \beta^2) (-1 +2\poisson) - 2 \Qq a^2\right] a_t^-
		\\\notag&- 8 r^3 [2 \Qq \poisson + \Pp (-1 + 2 \poisson)] \p_r a_t^-
		\\\notag&+ \left[4 \Pp r^4 (1-2\poisson) +4 \Qq r^2 (-2 r^2 \poisson + a^2)\right] \p_r^2 a_t^-
		\\\notag&+ 4 \Pp r^4 \beta (-1 +2 \poisson) \omega a_z^-
		\\\notag&-2 \ii \sqrt2 (\Pp + \Qq) r^3 (-1 + 2\poisson) \omega a_\sharp^-
		\\\notag&-2 \ii \sqrt 2 (\Pp + \Qq) r^4 (-1 + 2 \poisson) \omega \p_r a_\sharp^-
		\\\notag&-\ii \sqrt 2 \omega r \left[2 r^2 (\Pp (-3 + 6 \poisson) + \Qq (-1 + 6 \poisson)) + 4 \Qq a^2 \right] a_\flat^-
		\\&-2 \ii \sqrt 2 r^2 \omega \left[\Pp r^2 (-1 + 2\poisson) +\Qq r^2 (1+2\poisson) -2 \Qq a^2\right] \p_r a_\flat^-
	\Big\}
\,,
\\
\notag	\varepsilon \Sigma_t^{(-,2)} =& - n^2 \frac{\Bb}{4  r^2 a^2} \times
		\\\notag&\Big\{
		4\left[\Qq r^2 (1-2\poisson) + \Pp r^2 (-1 + r^2 \beta^2) (-1 +2\poisson) - 2 \Qq a^2\right] a_t^+
		\\\notag&- 8 r^3 [2 \Qq \poisson + \Pp (-1 + 2 \poisson)] \p_r a_t^+
		\\\notag&+ \left[4 \Pp r^4 (1-2\poisson) +4 \Qq r^2 (-2 r^2 \poisson + a^2)\right] \p_r^2 a_t^+
		\\\notag&+ 4 \Pp r^4 \beta (-1 +2 \poisson) \omega a_z^+
		\\\notag&-2 \ii \sqrt2 (\Pp + \Qq) r^3 (-1 + 2\poisson) \omega a_\flat^+
		\\\notag&-2 \ii \sqrt 2 (\Pp + \Qq) r^4 (-1 + 2 \poisson) \omega \p_r a_\flat^+
		\\\notag&-\ii \sqrt 2 \omega r \left[2 r^2 (\Pp (-3 + 6 \poisson) + \Qq (-1 + 6 \poisson)) + 4 \Qq a^2 \right] a_\sharp^+
		\\&-2 \ii \sqrt 2 r^2 \omega \left[\Pp r^2 (-1 + 2\poisson) +\Qq r^2 (1+2\poisson) -2 \Qq a^2\right] \p_r a_\sharp^+
	\Big\}
\,,
\\
	\varepsilon\Sigma_z^{(\pm,2)}  =& -\frac{  \Bb }{ a^2}  \Pp  (1-2\nu) n^4 \omega r^2 \left( \beta \mathring a_t^\mp + \omega \mathring a_z^\mp\right)
\,,
\\
	\varepsilon\Sigma_\sharp^{(+,2)}  =&  -\frac{ \Bb }{2  a^2} (1-2\nu) (\Qq +\Pp)  n^4 \omega r \left[2 r \omega \mathring a_\sharp^- - \ii \sqrt 2 \left(\mathring  a_t^- + r \p_r \mathring a_t^-\right)\right]
\,,
\\	\varepsilon\Sigma_\flat^{(-,2)} =&  -\frac{ \Bb }{2   a^2} (1-2\nu) (\Qq + \Pp)  n^4 \omega r \left[2 r \omega \mathring a_\flat^+ - \ii \sqrt 2 \left(\mathring  a_t^+ + r \p_r \mathring a_t^+\right)\right]
\,,
\\\notag
	\varepsilon\Sigma_\flat^{(+,2)} =&- \frac{\Bb }{ 2 r   a^2}  n^4 \omega \times
	\\\notag &\Big[ 2 (\Qq + \Pp) r^3 (1-2\nu)\omega \mathring a_\flat^- + 4\Qq r (a^2 - r^2) \omega \mathring a_\sharp^-
\\
\notag &- \ii \sqrt 2 \Big([2 \Qq a^2 - (3\Qq + \Pp) r^2  + 2 (\Qq + \Pp) r^2 \nu] \mathring a_t^- \\
& +r [2 \Qq a^2- (\Qq -\Pp)r^2  - 2 ( \Qq + \Pp)r^2 \nu] \p_r \mathring a_t^-\Big)\Big]
\,,
\\\notag
	\varepsilon\Sigma_\sharp^{(-,2)} =&- \frac{\Bb }{2 r  a^2} n^4  \omega \times
	\\\notag & \Big[ 2( \Qq + \Pp) r^3 (1-2\nu)\omega \mathring a_\sharp^+ +4 \Qq r (a^2 - r^2) \omega \mathring a_\flat^+
\\
\notag& - \ii \sqrt 2 \Big([2 \Qq a^2 -  (3 \Qq + \Pp) r^2 + 2 ( \Qq + \Pp) r^2 \nu] \mathring a_t^+
\\&+ r [2 \Qq a^2 -  ( \Qq -\Pp)r^2 - 2( \Qq  + \Pp)r^2 \nu] \p_r \mathring a_t^+\Big)\Big]
\,.
\end{align}
\end{subequations}
Further, for the correction terms arising for the interface conditions \eqref{eq:30I25} we find explicitly
\begin{eqnarray}\label{eq:30I25.2}
	\esGamma^{(\pm,0)} [\mathring a^\pm] =
	2 \Kk (\Pp + \Qq)
	\begin{pmatrix}
	 	0 \\
		0 \\
	 	0 \\
	  	0 \\
		\left\llbracket n^{4} [ \p_r \mathring a_t^\pm + \frac{\ii \omega}{\sqrt 2 } (\mathring a_\sharp^\pm + \mathring a_\flat^\pm)]\right\rrbracket \\
		0\\
		0 \\
		0\\
	\end{pmatrix}\,,
\end{eqnarray}
and
\begin{subequations}
\begin{align}\label{eq:30I25.3}
	\esGamma^{(+,2)}_5 [\mathring a^-] =
	-\frac{\Bb }{  \rho a^2} \Big\llbracket  &n^{4} \Big[\Qq (r^2 - a^2) \mathring a_t^- + r \left(r^2(-\Pp + 2 (\Pp + \Qq)\poisson) - \Qq a^2\right)\p_r \mathring a_t^- 
		\notag\\&+ \frac{\ii \omega}{ \sqrt 2} (\Pp + \Qq) r^3 (-1 + 2 \poisson)  \mathring a_\flat^- 
		\notag\\&+ \frac{\ii \omega}{ \sqrt 2} r (r^2 (-\Pp + \Qq + 2 \poisson (\Pp + \Qq))-2\Qq a^2) \mathring a_\sharp^- \Big] \Big\rrbracket \,,\\
		\esGamma^{(-,2)}_5 [\mathring a^+] =
	-\frac{\Bb }{  \rho a^2} \Big\llbracket  &n^{4} \Big[\Qq (r^2 - a^2) \mathring a_t^+ + r (r^2(-\Pp + 2 (\Pp + \Qq)\poisson) - \Qq a^2)\p_r \mathring a_t^+
		\notag\\&+ \frac{\ii \omega}{ \sqrt 2} (\Pp + \Qq) r^3 (-1 + 2 \poisson)  \mathring a_\sharp^+
		\notag\\&+ \frac{\ii \omega}{ \sqrt 2} r (r^2 (-\Pp + \Qq + 2 \poisson (\Pp + \Qq))-2\Qq a^2) \mathring a_\flat^+ \Big] \Big\rrbracket \,,
\end{align}
\end{subequations}
with the remaining components of $\esGamma^{(\pm,2)}$ vanishing identically.

Based on these source terms, equation \eqref{24VI24.2} determining the transport matrix for the Jones vector can be evaluated numerically by inserting the relevant Bessel functions and integrating.
We find that the matrix $\mathcal M$ in \eqref{24VI24.2} has the form
\begin{align}
	\mathcal M
		&=
		\begin{pmatrix}
				\xi_p \Kk
			&	\xi_b \Bb
			\\
				\xi_b \Bb
			&	\xi_p \Kk
		\end{pmatrix}\,.
\end{align}
The corresponding matrix $\hat{\mathcal M}$ entering the propagation law \eqref{eq:Jones transport cartesian} in Cartesian coordinates thus takes the form
\begin{align}
	\hat{\mathcal M}
		&=
		\xi_p \Kk
		\begin{pmatrix}
			1&	\hphantom{-}0
			\\
			0
			&	\hphantom{-}1
		\end{pmatrix}
		+ \xi_b \Bb
		\begin{pmatrix}
			1	
			&	\hphantom{-}0
			\\
			0
			&	-1
		\end{pmatrix}\,.
\end{align}
Here, $\xi_p$, $\xi_b$ are coefficients that depend only on the fiber core radius, its unperturbed refractive indices, and the optical frequency of the fiber mode. Numerical results are presented in Sect.~\ref{sec:numerics}.

Since the unperturbed solutions described in Sect.~\ref{sec:unperturbed} are confined to a small region around the fiber’s axis, with the modes decaying exponentially in the cladding, it is natural to consider an approximate form of the above expressions that is valid for $r \ll a$.
Using this \emph{confinement approximation}, the strain components take the form
\begin{subequations}\label{eq:15I25.1}
\begin{align}
	\begin{split}
		u_{rr} ={}
			& \shearmod^{-1} \Big[ (1-2\poisson)  d_0 - \shearmod \poisson  \alpham +  \shearmod(1+ \poisson) \alphat \deltaT
			+   b_2 \cos(2 \theta)
			 \Big]\,,
	\end{split}
	\\
	\begin{split}
		u_{r\theta} ={}
			&-\shearmod^{-1}   b_2\sin (2 \theta)\,,
	\end{split}
	\\
	\begin{split}
		u_{\theta\theta} ={}
			&  \shearmod^{-1}  \Big[  (1-2\poisson)  d_0  - \shearmod \poisson \alpham + \shearmod(1+ \poisson) \alphat \deltaT
			-   b_2\cos(2\theta)
			 \Big]\,.
	\end{split}
\end{align}
\end{subequations}
Thus, the source terms for $\Delta m = 0$ are identical to \eqref{eq:16XII24.2a}--\eqref{eq:16XII24.2d} and \eqref{eq:30I25.2}, while $\Sigma_b^{(\pm,2)}$ and $\esGamma^{(\pm,2)}$ simplify significantly to
\begin{subequations}
\begin{align}
\varepsilon \Sigma_t^{(+,2)} &= \frac{\Qq n^2}{ r^2 } \Bb  \left\{ 2 a_t^- -r^2 \p_r^2 a_t^- +\ii \sqrt 2 \omega r \left(a_\sharp^- - r \p_r a_\sharp^-\right) \right\}
\,,
\\
\varepsilon \Sigma_t^{(-,2)} &=\frac{\Qq n^2}{ r^2}  \Bb  \left\{ 2 a_t^+ - r^2 \p_r^2 a_t^+ + \ii \sqrt 2 \omega r \left(a_\flat^+ - r \p_r a_\flat^+\right) \right\}
\,,
\\
	\varepsilon\Sigma_z^{(\pm,2)}  &= 0
\,,
\\
	\varepsilon\Sigma_\sharp^{(+,2)}  &=  0
\,,
\\	\varepsilon\Sigma_\flat^{(-,2)} &=  0
\,,
\\
	\varepsilon\Sigma_\flat^{(+,2)} &= -\frac{ \Qq n^4 \omega}{r}   \Bb \left[ 2 r \omega a_\sharp^- - \ii \sqrt 2 (a_t^- + r \p_r a_t^-)\right]
\,,
\\
	\varepsilon\Sigma_\sharp^{(-,2)} &=-\frac{ \Qq n^4 \omega}{r }  \Bb \left[ 2 r \omega a_\flat^+ - \ii \sqrt 2 (a_t^+ + r \p_r a_t^+)\right]
\,,
\end{align}
\end{subequations}
and
\begin{eqnarray}
	\esGamma^{(\pm,2)} [\mathring a^\mp] =
	-\frac{ \Qq  }{\rho } \Bb
	\begin{pmatrix}
	 	0 \\
		0 \\
	 	0 \\
	  	0 \\
		\left\llbracket n^{4} [\mathring a_t^\mp + r \p_r \mathring a_t^\mp + \ii \sqrt{2} r \omega \mathring a_\flat^\mp]\right\rrbracket \\
		0\\
		0 \\
		0\\
	\end{pmatrix}\,.
\end{eqnarray}
The numerical results produced in Table~\ref{tab:GRAVITES} show good agreement between the confinement approximation and the full solution.

\subsection{Numerical results}\label{sec:numerics}

The sections \ref{sec:displacement} and \ref{sect:photoelasticity} provide explicit expressions for the $\Sigma$-
and $\esGamma$-terms
that enter the matrix $\mathcal M$, defined in Eq.~\eqref{24VI24.2}, which determines the propagation law of light polarization according to Eqs.~\eqref{eq:Jones transport complex} and \eqref{eq:Jones transport cartesian}.
Here we present  numerical values based on the material properties of silica oxide glass given in Table~\ref{tab:SiO2} and the experimental parameters of GRAVITES given in Table~\ref{tab:parGRAVITES}. The values for $\delta \grav$, $\delta \mathfrak{p}$ and $\deltaT$ correspond to  a height difference of $1\,\mathrm{m}$ for an experiment without thermal insulation and pressure control. Since the response of the system is linear in the environmental parameters,  these numbers  can directly be used to determine the constraints on the experimental setup so that environmental fluctuations do not overwhelm the signal of interest.

The numerical values determined in this work are juxtaposed in Table~\ref{tab:GRAVITES} with the results of Ref.~\cite{BCS24}, where the change of length  of the optical fiber has been determined.

In Figures~\ref{fig:birefringence} and \ref{fig:phase shifts} we provide plots showing how a change in fiber parameters affects these values.
Both for phase and birefringence the effect is essentially due to the photoelastic material response, as the contribution arising from the core-cladding interface deformation is smaller by four orders of magnitude. Additionally, this effect is practically independent of the fiber’s material indices $n_1$ and $n_2$.  
\begin{table}[h!]
	\centering
	\begin{tabular}{r|l}
			Fiber outer radius $a$
		&	\SI{62.5}{\micro\meter}
		\\
			Fiber core radius $\rho$
		&	\SI{4.1}{\micro\meter}
		\\
			Core refractive index $n_1$
		&	1.4712
		\\
			Cladding refractive index $n_2$
		&	1.4659
		\\
			Variation in grav.\ acceleration, $\delta\grav$
		&	\SI{3}{\micro\meter\per\square\second}
		\\
			Variation in pressure, $\delta \mathfrak p$
		&	\SI{10}{\pascal}
		\\
			Variation in temperature, $\deltaT$
		&	\SI{e-2}{\kelvin}
		\\
			Fiber length $L$
		&	\SI{e5}{\meter}
		\\
			Propagation constant $\beta$
		&	\SI{6e6}{\per\meter}
		\\
			Wavelength $\lambda$
		&	\SI{1.55}{\micro\meter}
	\end{tabular}
\caption{Parameters for the GRAVITES experiment \cite{Hilweg:2017ioz}.}
\label{tab:parGRAVITES}
\end{table}
\begin{table}[h!]
\centering
\begin{tabular}{@{}lSSS[table-format=3.1e2]S[table-format=3.1e2]@{}}
	\toprule
	Gravitational phase shift
	&	\multicolumn{4}{c}{\num{-6.52e-5}}
	\\\midrule
	\multirow{2}{*}{Systematic effects}
	&	\multicolumn{1}{c}{Temperature}
	&	\multicolumn{1}{c}{Pressure}
	&	\multicolumn{2}{c}{Gravity gradient}
	\\
	\cmidrule(lr){2-2}
	\cmidrule(lr){3-3}
	\cmidrule(ll){4-5}
	&	\multicolumn{1}{c}{\small Phase}
	&	\multicolumn{1}{c}{\small Phase}
	&	\multicolumn{1}{c}{\small Phase}
	&	\multicolumn{1}{c}{\small Birefringence}
	\\\midrule
	Longitudinal expansion \cite{BCS24}
	&	3240
	&	28
	&	7.0e-7
	&
	\\
	Geometric deformation
	&	8.7
	&	-0.15
	&	-3.0e-9
	&	2.42e-11
	\\
	Photoelasticity
	&	3190
	&	-53
	&	-1.1e-6
	&	-1.27e-6
	\\
	Photoelasticity confinement approx.
	&	3190
	&	-53
	&	-1.1e-6
	&	-1.28e-6
	\\\bottomrule
\end{tabular}
\caption{Estimates for the phases, in radians, arising in the GRAVITES experiment described in \cite{Hilweg:2017ioz}.
The phase shifts arising from the longitudinal fiber expansion were computed in Ref.~\cite{BCS24} The first column differs due to a correction of the linear thermal expansion coefficient $\alpha$; the remaining shifts are determined
 in this work.
Temperature and pressure  differences over a vertical separation of \SI{1}{\meter} give rise to phase shifts via elastic deformations of the fiber and induced photoelastic effects;  gravity gradients induce inhomogeneities and thus give rise to both phase perturbations and birefringence effects.
Confinement approximation refers to \eqref{eq:15I25.1} wherein the material strain components are expanded for $r\ll a$.
}
\label{tab:GRAVITES}
\end{table}

\section{Conclusion}

We solved Maxwell's equations with small source terms in an optical fiber background to first order in a perturbation scheme. We gave a general expression of the propagation of the Jones vector, which encodes changes in phase and birefringence, along the axis of the fiber to first order in perturbation theory -- \eqref{eq:Jones transport cartesian}.

Using results derived in \cite{BCS24} for elastic deformations of cylinders we give formulas for the explicit source terms corresponding to interface deformations of the optical fiber and stress induces changes in the optical properties of the waveguide.

As a practical application of these findings we compute the expected changes in phase and birefringence due to elasticity effects for a range of fiber parameters in Figures~\ref{fig:birefringence}--\ref{fig:phase shifts} and concretely for the GRAVITES experiment \cite{Hilweg:2017ioz} in Table~\ref{tab:GRAVITES}.
Concerning the effect of temperature changes we note that the results here are restricted to elastic contributions and do not encompass the full thermo-optic effect. Experimentally it is known that the longitudinal thermal expansion contributes about $5\%$ to the total temperature dependence of the optical phase in fused silica fibers  \cite{10.1117/12.2193091}. Our calculations show that the contribution of photoelasticity is on the same order; this differs from material to material \cite{Waxler1973TheEO}. 

\FloatBarrier
\section*{Acknowledgments}
E.S.\ was supported in part by the Austrian Science Fund (FWF), Project P34274.
The research of P.T.C.\ and T.B.M.\ was funded by the European Union (\textsc{erc}, \textsc{gravites}, project no.~101071779).
Views and opinions expressed are however those of the authors only and do not necessarily reflect those of the European Union or the European Research Council Executive Agency. Neither the European Union nor the granting authority can be held responsible for them.

\end{document}